\documentclass[reprint,aps,twocolumn,groupedaddress]{revtex4-1}
\usepackage{color}
\usepackage{graphicx}
\usepackage{enumerate}
\usepackage{amsfonts,amssymb}
\usepackage{amsmath}
\usepackage{braket}
\usepackage{verbatim}
\usepackage{bbold}
\usepackage{svg}

\makeatletter
\def\maketitle{
\@author@finish
\title@column\titleblock@produce
\suppressfloats[t]}
\makeatother

\begin{document}

\title{Entanglement in the quantum phases of an unfrustrated Rydberg atom array}
\author{Matthew J. O'Rourke}
\affiliation{\footnotesize{Division of Chemistry and Chemical Engineering,
California Institute of Technology, Pasadena, CA 91125, USA}}
\author{Garnet Kin-Lic Chan}
\affiliation{\footnotesize{Division of Chemistry and Chemical Engineering,
California Institute of Technology, Pasadena, CA 91125, USA}}
\bigskip
\date{\today}

\begin{abstract}
We report on the ground state phase diagram of interacting 
Rydberg atoms in the unfrustrated square lattice array. 
Using new tensor network algorithms, we scale to large systems in two dimensions while including all long-range interactions, revealing the phases in the bulk and their analogs in accessible finite arrays. We find a greatly altered phase diagram from earlier numerical and experimental studies, and in particular, we uncover an emergent entangled quantum nematic phase that appears in the absence of frustration. Broadly our results yield a conceptual guide for future experiments, while our techniques provide a blueprint for converging numerical studies in other lattices.
\end{abstract}
\maketitle

\section{Introduction}
Rydberg atom arrays, where cold atoms are trapped
in an optical lattice and interact 
via excitation into Rydberg states~\cite{bloch2008,saffman2016}, have generated interest for quantum information processing and to realize exotic many-body states~\cite{endres2016atom,barredo2016atom,
lee2017defect,brown2019gray,ohl2019defect,barredo2018synthetic,kumar2018sorting,scholl2020,pichler2018quantum,pichler2018computational,bernien2017probing,keesling2019quantum,de2019observation,samajdar2021quantum,verresen2020prediction,samajdar2021quantum}.
A recent experiment~\cite{ebadi2020}, backed by numerical
studies~\cite{lukin2020,felser2021}, has suggested a richness
in 2D Rydberg atom array ground states on a square lattice. However, although the observed, non-disordered, phases are not all classical crystals, they contain little entanglement~\cite{lukin2020}. Thus it remains unclear whether such arrays realize non-trivial entangled quantum ground-states on simple lattices. A confounding feature suggested more recently~\cite{verresen2020prediction},
is that the long-range tails of the interactions greatly affect the phases, complicating the accurate numerical determination of bulk behaviour. Here, we describe new numerical techniques that greatly reduce finite size effects, allowing us to confidently converge the bulk phase diagram. We also showcase techniques that address large finite two-dimensional lattices realized in experiments, while incorporating all long-range interactions. Unexpectedly, we derive quite different physics from our simulations compared to both previous theoretical and experimental analyses -- including the emergence of a non-trivial, entangled nematic phase, even on the unfrustrated square lattice array.

The Rydberg atom array Hamiltonian is
\begin{equation}
\hat{H} =  \sum_{i=1}^N \left[ \frac{\Omega}{2} \hat{\sigma}_i^x - \delta \hat{n}_i \right] + 
    \frac{1}{2} \sum_{i \neq j} \frac{V}{(|\vec{r}_i - \vec{r}_j|/a)^6} \hat{n}_i \hat{n}_j.
\label{eqn:Hamil}
\end{equation}
Here $\hat{\sigma}_i^x = 
\ket{0_i}\bra{1_i} + \ket{1_i}\bra{0_i}$ and $\hat{n}_i = \ket{1_i}\bra{1_i}$ ($\{ \ket{0_i}, \ket{1_i}\}$  denote ground and Rydberg states of atom $i$). 
 $a$ is lattice spacing,
$\Omega$ labels Rabi frequency, and $\delta$ describes laser detuning.  $V$ parameterizes the interaction strength between excitations. This  can  be re-expressed in terms of the Rydberg blockade radius $R_b$,  with 
$V / (R_b/a)^6 \equiv \Omega$. 
We study the square lattice in
 units  $a = \Omega = 1$~\cite{lukin2020}, yielding two free 
parameters $\delta$ and $R_b$.

The ground states of this Hamiltonian are simply understood in two limits. 
For $\delta / \Omega \gg 1$, $R_b \neq 0$, 
the system is classical and one obtains
classical crystals of Rydberg excitations~\cite{bak1982commensurate,bak1982one,fendley2004,science2015}
whose spatial density is
set by the competition between $\delta$ and $R_b$. 
For $\delta / \Omega \ll 1$, $R_b \neq 0$, Rydberg
excitations are disfavored and the solutions are dominated by Rabi oscillations, leading to a trivial ``disordered'' phase~\cite{weimer2010,lauchli2019,lukin2020}.
In between these limits, it is known in 1D that
no other density-ordered ground states exist besides the classical-looking crystals, with 
 a Luttinger liquid appearing on the boundary between ordered and disordered phases~\cite{lauchli2019}. 

In 2D, however, the picture is quite different. An initial study~\cite{lukin2020} using
the density matrix renormalization
group (DMRG)~\cite{2Ddmrg2012,white1992dmrg,white1993dmrg,schollwock2011density}
found additional quantum crystalline (or ``density-ordered'') phases, where the local excitation density is not close to $0$ or $1$.
 A recent experiment on a 256 programmable atom array has realized such phases~\cite{ebadi2020}.  However, as also discussed there,
the density-ordered phases are unentangled quantum mean-field phases, and thus not very interesting. In addition, more recent numerical results~\cite{verresen2020prediction} highlight the sensitivity of the physics to the tails of the Rydberg interaction and finite size effects. Thus, whether Rydberg atom arrays on a simple unfrustrated lattice -- such as the square lattice -- support interesting quantum ground-states, remains an open question.

\begin{figure*}
    \centering
    %\includegraphics[width=\textwidth]{figs/diagrams_final_12_2.png}
    %\includesvg[width=\textwidth]{figs/diagrams_final_12_2.svg}
    \includegraphics[width=\textwidth]{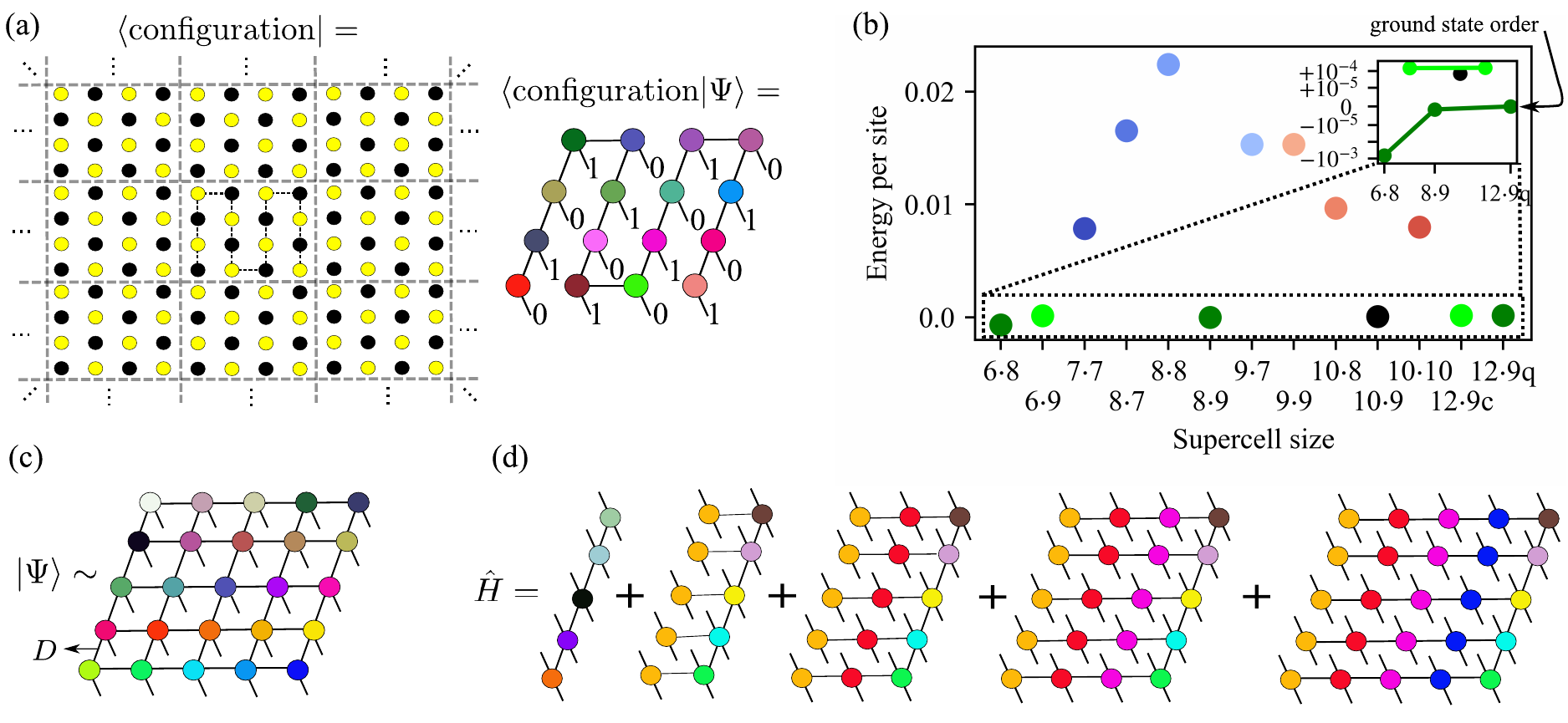}
    \caption{\textbf{Numerical methods and strategy.} 
    (a) A schematic representation of $\Gamma$-point DMRG. A single infinite bulk configuration
    is given by periodic images of the central supercell configuration. 
    The wavefunction coefficient for this infinite configuration is given by
    the contraction of a snake MPS, which is defined only within a single supercell.
    (b) By widely varying the size of the supercell, $\Gamma$-point DMRG obtains many different
    ground states. Identifying all accessible supercells which give the same ground state order
    (shown with identically colored points), we can ensure that all competing low-energy states
    are well converged w.r.t. finite size effects, and thus properly identify
    the true ground state (inset shows ground-state order (dark green) 
    converged w.r.t. supercell size, separated from other 
    low-energy orders by $10^{-4}$ energy units).
    (c) A PEPS wavefunction 
    ansatz with bond dimension $D$ for a finite system. Each tensor is a different color because
    they can all be unique.
    (d) A simplified diagrammatic representation of the
    long-range Hamiltonian construction for PEPS in Ref.~\cite{gmpos}. All terms in the
    Hamiltonian are accounted for by a sum of $L_x$ comb tensor network operators.
    Tensors of the same color are identical.
     }
     \label{fig:diagrams}
\end{figure*}

Here, we resolve these questions through high-fidelity numerical simulations.
To do so, we employ variational tensor network methods.
Tensor networks have led to breakthroughs in the understanding of 2D quantum many-body problems~\cite{zheng2017stripe}, and we rely on two new techniques that address specific complexities of simulating interactions in Rydberg atom arrays. The first we term $\Gamma$-point DMRG, which captures interactions out to \emph{infinite} range, while employing a traditional finite system two-dimensional DMRG methodology~\cite{2Ddmrg2012}, removing interaction truncations and boundary effects present in earlier studies~\cite{lukin2020,verresen2020prediction,samajdar2021quantum,felser2021}.
This allows us to controllably converge the bulk phase diagram.
The second is a representation of long-range interactions~\cite{gmpos} compatible  
with projected entangled pair states 
(PEPS)~\cite{verstraete2004renormalization,verstraete2006criticality,nishino1996corner,orus2014practical}. 
With this, we use PEPS to find the ground states of a Hamiltonian
with long-range interactions for the first time, and 
specifically here, model the states of finite arrays of large widths as used in experiment. Both techniques can be used for more faithful simulations of Rydberg atoms in other settings. We first describe the new numerical methods, 
before turning to the bulk and finite-size phase behavior of square lattice Rydberg arrays and the 
question of entangled quantum phases.

\section{Numerical strategy and techniques}
\label{sec:strategy}

\subsection{Bulk simulations and $\Gamma$-point DMRG}

\label{sec:gammaDMRG}

A challenge in simulating Rydberg atom arrays 
is the long-range tails of the interaction. Because itinerancy
only arises indirectly as 
an effective energy scale~\cite{weimer2010},
the main finite size effects arise from interactions.
Many previous studies have employed a cylindrical DMRG geometry common
in 2D DMRG studies~\cite{2Ddmrg2012}. However, there the interaction is necessarily truncated to the cylinder half-width, while along the open direction, edge atoms 
experience different interactions than in the bulk; both choices produce strong finite size effects.

To avoid these problems, we perform 2D DMRG calculations in a Bloch basis. The
resulting $\Gamma$-point 2D DMRG formally models an infinite lattice (Fig.~\ref{fig:diagrams}a) with a wavefunction constrained by the supercell, and periodic boundary conditions in both directions. This  differs from using a periodic matrix product state (MPS) as periodicity is enforced by the Bloch basis rather than the MPS, and the underlying 2D DMRG can be carried out using the conventional snake path. It is also different from a cylindrical/toroidal geometry, which are both finite; here the lattice remains infinite. The Bloch basis interactions enter
as an infinite lattice sum over lattice vectors $R_l = (n \cdot L_x, m \cdot L_y)$;
$n,m \in \mathbb{Z}$; $i \neq j + R_l$
\begin{equation}
\hat{H} = \sum_{i} \left[ \frac{1}{2} \hat{\sigma}_i^x - \delta \hat{n}_i \right]
+ \frac{1}{2} \sum_{i \neq j + R_l, R_l} \frac{R_b^6}{|\vec{r}_i - \vec{r}_{j+R_l}|^6} \hat{n}_i \hat{n}_j.
\end{equation}
where $L_x$, $L_y$ are the supercell side lengths.

The only finite size parameter is the supercell size $L_x \times L_y$. We thus perform exhaustive scans over $L_x, L_y$. However, because no interactions are truncated and there are no edge effects even in the smallest cells, finite size effects converge very rapidly (much more quickly than using a cylinder or torus). Using different supercell sizes with up to 108 sites we converge the energy per site to better than $10^{-5}$, compared to the smallest
energy density difference we observe between competing phases of $\sim 10^{-4}$ (see Fig.~\ref{fig:diagrams}b and supplementary information~\cite{supp}).

\subsection{Finite simulations and PEPS with long-range interactions}

\label{sec:lrPEPS}

To simulate ground-states of finite arrays, we consider finite systems (with open boundaries) 
of sizes $9\times 9$ up to $16 \times 16$ atoms. This 
resembles capabilities of near-term 
experiments~\cite{scholl2020,ebadi2020}. The width of the largest arrays challenges
what can be confidently described with MPS and DMRG for more entangled states.
Consequently, we employ PEPS wavefunctions which capture 
area law entanglement in 2D, and can thus be scaled to very wide
arrays (Fig.~\ref{fig:diagrams}c). Together with DMRG calculations on moderate width finite lattices, the two methods provide  complementary approaches to competing phases and consistency between the two provides strong confirmation. However, PEPS are usually combined with short-range Hamiltonians.
We now discuss a way to combine long-range Hamiltonians efficiently with PEPS without truncations.

For this, we rely on the
representation we introduced in Ref.~\cite{gmpos}. This encodes the long-range Hamiltonian
%sum 
%over a large number of pair interactions
%(more than $32000$ on a $16\times 16$ lattice) by
as a sum of ``comb'' tensor network operators (Fig.~\ref{fig:diagrams}d). As discussed in Ref.~\cite{gmpos}, arbitrary isotropic interactions
can be efficiently represented in this form, which mimics the desired potential via a sum of Gaussians, i.e. $\frac{1}{r^6} = \sum_{k=1}^{k_\text{max}} c_k e^{-b_k r^2}$ (where $k_\text{max}\sim 7$ for the desired accuracy in this work). The combs can be efficiently contracted much more cheaply than using a general tensor network operator.

While Ref.~\cite{gmpos} described the Hamiltonian encoding, here we must also find the ground-state. We variationally minimize $\langle \Psi|\hat{H}|\Psi\rangle$ using  automatic differentiation~\cite{autodiffTN}. Combined with the comb-based energy evaluation,
this allows for both the PEPS energy and gradient to be evaluated with a cost linear in lattice size. (Stably converging the PEPS optimization involves some challenges. Further details in~\cite{supp}).

\section{Bulk phases}
\label{sec:pbc}

\begin{figure*}
    \centering
    \includegraphics[width=\textwidth]{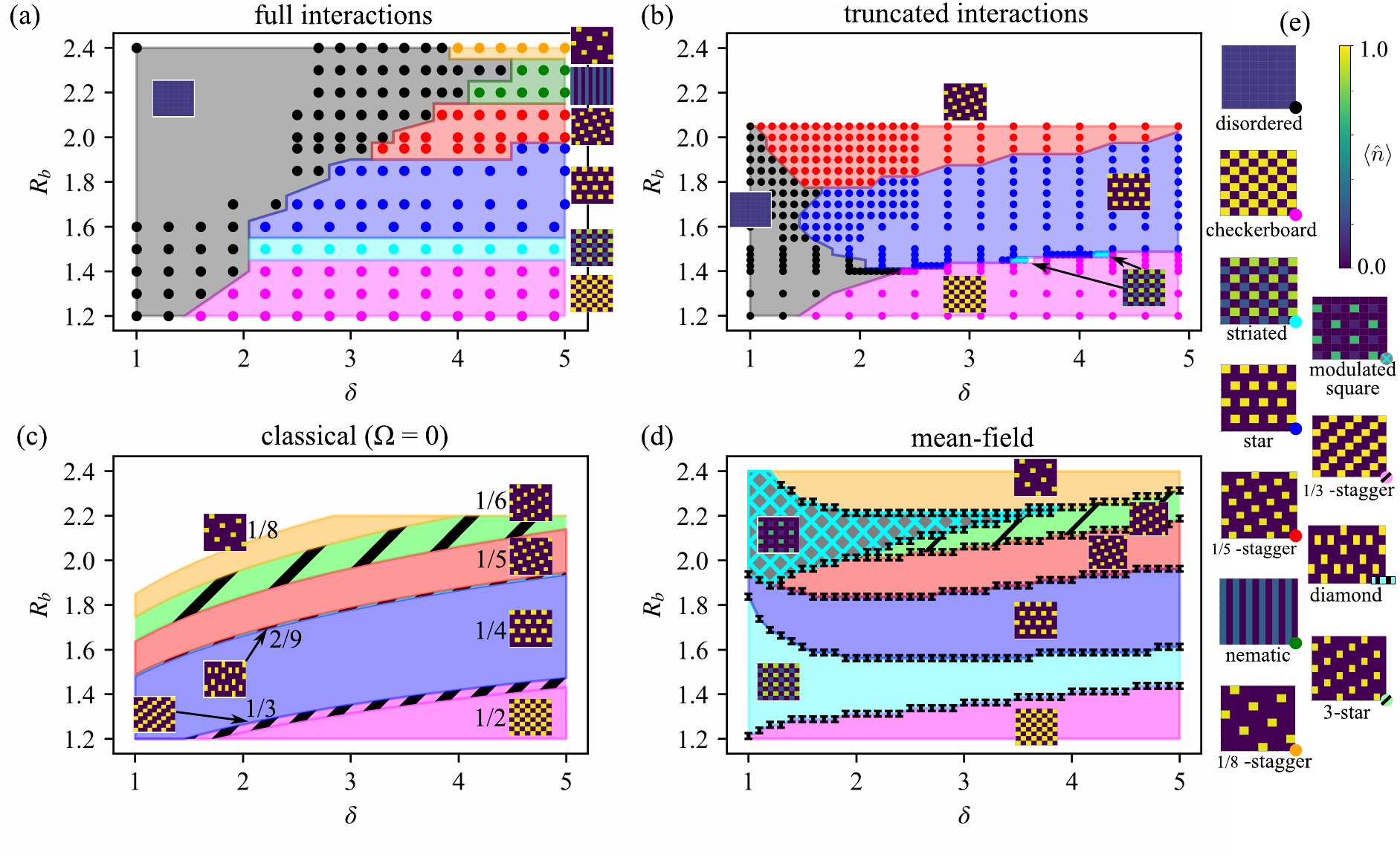}
    \caption{\textbf{Phase diagrams of the bulk system under various assumptions.}
    The color of a dot/region identifies the ground state order. The density profiles for each 
    color are given in (e) and shown near each phase domain. 
    (a) The phase diagram given by $\Gamma$-point DMRG including all long-range interactions.
    (b) The phase diagram from $\Gamma$-point DMRG when interactions are truncated to 0
    beyond a distance of $|\vec{r}_i - \vec{r}_j| = 2$.
    (c) The classical phase diagram (when all sites are either fully occupied or
    empty) including all long-range interactions.
    (d) The mean-field phase diagram, including all long-range interactions.
    Error bars display the uncertainty of the computed phase boundaries.
    (e) Representative density profiles for all phases in (a)-(d), identified by 
    the colored dot in each lower right corner. All profiles have $\Gamma$-point 
    boundary conditions on all edges. 
    In (a)-(b) dots denote computed
    data, while shading is a guide for the eye. (c),(d) are computed with very fine
    resolution/analytically, thus no dots are shown.}
     \label{fig:pbc}
\end{figure*}

\noindent \textbf{Summary of the phase diagram}. Fig.~\ref{fig:pbc}a shows the bulk phase diagram from $\Gamma$-point DMRG with infinite-range interactions. We first discuss the orders identified by their density profiles (orders of some phase 
transitions are briefly discussed in~\cite{supp}). 
Where we observe the same phases as in earlier work~\cite{lukin2020} we use the same names, although there are very substantial differences with earlier phase diagrams.

\noindent \underline{$R_b < 1.8$}. 
With {weaker} interactions, the ground states progress through densely-packed, density-ordered phases starting from 
checkerboard (pink) ($R_b\sim 1.2$), to striated (cyan) ($R_b\sim 1.5$), to star (blue) ($R_b\sim 1.6$).
 While the checkerboard and star phases are classical-like crystals, the striated state is a density-ordered quantum phase, seen previously~\cite{lukin2020}.

\noindent \underline{$R_b>1.8$}. 
Here, the phases look very different from earlier work, which truncated the interactions~\cite{lukin2020}.
Ordered ground states start with the $\frac{1}{5}$-``staggered'' phase (red) ($R_b \sim 1.95$), 
then progress to a ``nematic'' phase (dark green) ($R_b \sim 2.2$) and 
the $\frac{1}{8}$-``staggered'' phase (gold) ($R_b\sim 2.4$).
There is also a small
region at larger $\delta$ (not shown) where the nematic phase and a ``3-star'' classical-like 
crystal appear to be essentially degenerate, with an energy difference per site of $\Delta e < 3\cdot10^{-5}$ (see~\cite{supp}).

\noindent{\textbf{Effects of interactions}}.
In Fig.~\ref{fig:pbc}b we show the phase diagram computed using $\Gamma$-point DMRG with
interactions truncated to distance $2$. This approximation resembles earlier numerical
studies~\cite{lukin2020}, but here bulk boundary conditions are enforced by the Bloch 
basis, rather than cylindrical DMRG. Comparing Figs.~\ref{fig:pbc}a,b we see the disordered
and striated phases are greatly stabilized using the full interaction, 
and new longer-range orders are stabilized at larger $R_b$.
Comparing Fig.~\ref{fig:pbc}b and Ref.~\cite{lukin2020}, we see that having
all atoms interact on an equal footing (via the Bloch basis) destroys the quantum
ordered phases seen in~\cite{lukin2020}.

\noindent \textbf{Classical, mean-field, and entangled phases}.
\label{sec:pbc_quantum} 
Without the Rabi term $\Omega$, one would obtain classical Rydberg crystals without a disordered phase. 
Fig.~\ref{fig:pbc}c shows the classical phase diagram.
For the $\delta$ values here, the 1D classical phase diagram has sizable regions of stability for all accessible unit fraction densities~\cite{bak1982one,lauchli2019}. However, the connectivity of the square lattice in 2D changes this. For example,
only a tiny part of the phase diagram supports
a $\frac{1}{3}$-density crystal, and we do not find a stable $\frac{1}{7}$-density crystal within unit cell sizes of up to $10 \times 10$.  
All ordered quantum phases in Fig.~\ref{fig:pbc}a appear as classical phases except for the striated and nematic phases, while there are small regions of classical phases at densities $\frac{1}{3}$ and 
$\frac{2}{9}$ with no quantum counterpart. 
The striated and nematic phases emerge near the $\frac{1}{3}$ and $\frac{1}{7}$ density 
gaps respectively, however the nematic phase also supersedes the large region of the $\frac{1}{6}$ density ``3-star" crystal.

\begin{figure*}
    \centering
    %\includegraphics[width=\textwidth]{figs/correlation_fns_12_2.png}
    %\includesvg[width=\textwidth]{figs/correlation_fns_12_2.svg}
    \includegraphics[width=\textwidth]{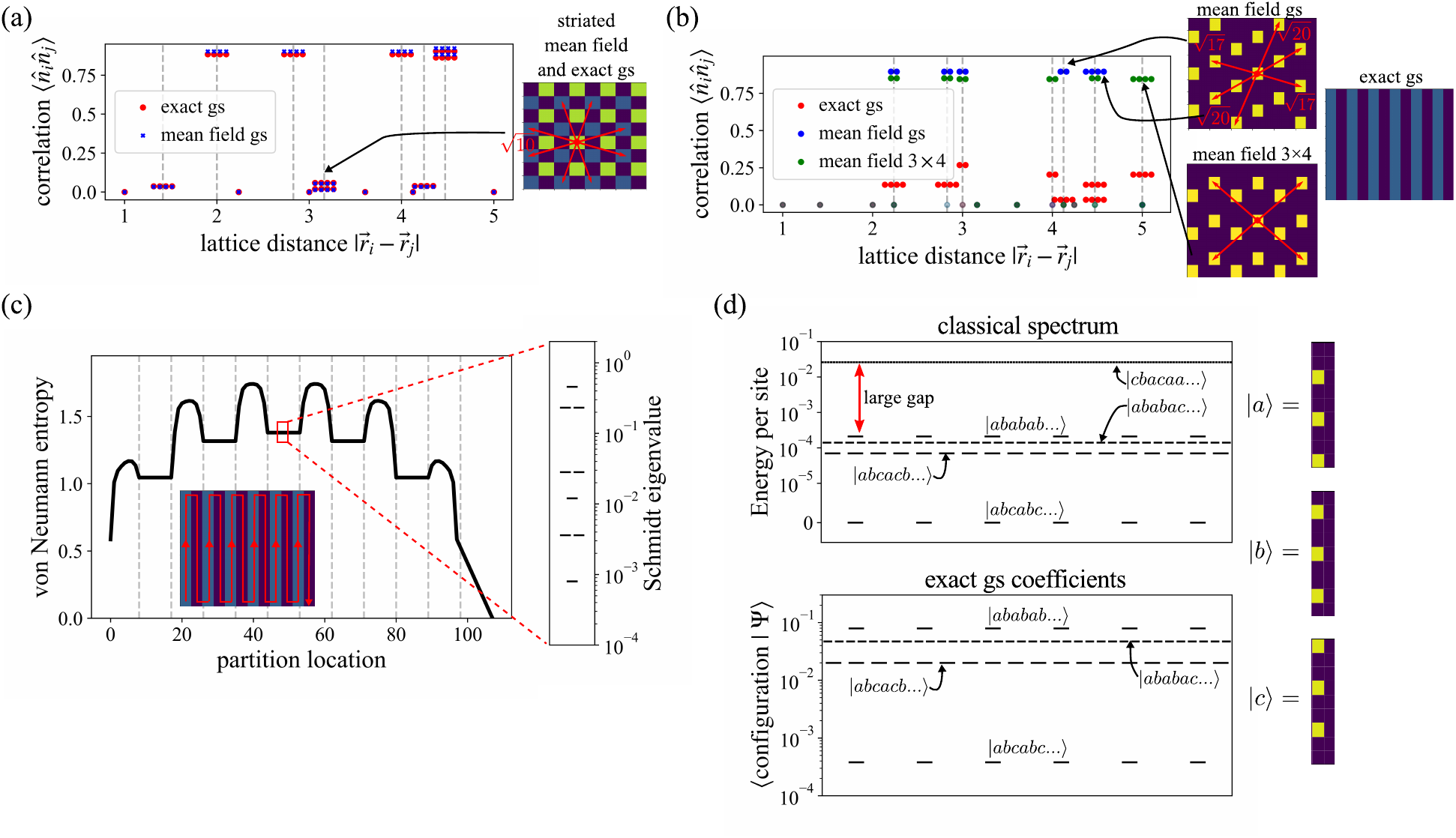}
    \caption{\textbf{Mean-field striated versus entangled nematic phase.}
    (a) Density-density correlation functions of the mean-field and exact striated ground state, both at $(\delta, R_b) = (3.1, 1.5)$; these agree, confirming the mean-field nature of the striated phase.
    (b) Density-density correlation functions for the entangled nematic phase ground state and two different mean-field ground states (from a $6 \times 3$ unit cell and a $3 \times 4$ unit cell) at $(\delta, R_b) = (5.0, 2.3)$. In (a)-(b), 2-fold/4-fold degeneracy of a peak is indicated by 2/4 horizontal dots distributed around the proper distance coordinate.
    8-fold degeneracy in (a) is shown as two rows of 4 dots. The non-mean-field (entangled) character of the nematic phase is evident.
    (c) Bipartite entanglement entropy for each possible bipartition of the
    $12 \times 9$ supercell nematic ground state. One inset shows the ``path''
    that the ``partition location'' axis follows through the supercell MPS, while the other
    shows the entanglement spectrum at a central cut.  
    (d) Structure of the nematic state in terms of classical configurations constructed via compositions of 3 individual column states $\ket{a}, \ket{b}, \ket{c}$. In the classical limit,
    there are 4
    distinct sets of low-energy configurations, all characterized by the absence of     adjacent columns in the same state (e.g. $\ket{aa...}$) and large degeneracies due to 
    permutational symmetry between $\ket{a}$, $\ket{b}$, and $\ket{c}$. The lowest in energy is 6-fold
    degenerate, corresponding to the 3-star state. However, in the quantum nematic state the configurations that are slightly higher in energy have much larger wavefunction coefficients.
    The most relevant classical states in the wavefunction are those with the greatest number     of possible single ``column hops'' (e.g. $a \rightarrow b$)      without introducing unfavorable states like $\ket{aa...}$,
    revealing the role of itinerancy in the nematic phase.
    }
    \label{fig:nematic}
\end{figure*}

Ref.~\cite{ebadi2020} suggested that quantum density-ordered phases are
qualitatively mean-field states of the form $\prod_i \alpha_i |0_i\rangle + \sqrt{1-|\alpha_i|^2}|1_i\rangle$. 
Fig.~\ref{fig:pbc}d  shows the mean-field phase diagram. The disordered phase does not appear, as it emerges from defect hopping and cannot be described without some entanglement~\cite{weimer2010}. 
The mean-field phase diagram contains features of both the classical and quantum phase diagrams. The striated quantum phase indeed appears as a mean-field state, 
confirmed by the match between the mean-field and exact correlation functions (Fig.~\ref{fig:nematic}a). 
However, the nematic phase does \emph{not} appear, and 
in its place is the same $\frac{1}{6}$-density crystal stabilized in the classical phase diagram.
The nematic phase thus emerges as an example of the non-trivial entangled ground-state we are seeking.

\noindent\textbf{Nature of the entangled nematic phase}. Fig.~\ref{fig:nematic}b shows the density correlation function of the nematic phase, which does not display mean-field character. The
bi-partite entanglement entropy and
entanglement spectrum are shown in Fig.~\ref{fig:nematic}c. Importantly, 
the entanglement spectrum carries 3 large Schmidt values across every cut along the DMRG snake MPS, showing the state is fully entangled across the supercell, and well approximated by an MPS of bond dimension 3.

To reveal the phase structure, Fig.~\ref{fig:nematic}d shows the lowest energy classical states in the same region of the phase diagram. Due to the Rydberg blockade radius ($R_b=2.3$), excitations are spaced by 3 units within a column, giving 3 column configurations  $|a\rangle$, $|b\rangle$, $|c\rangle$. Column-column interactions, however, prevent adjacent columns from being in the same configuration (which would have excitations separated by 2 units); this is an adjacent-column constraint. The lowest classical state is the crystal $|abc\ldots\rangle$ (the 3-star phase) and its 6-fold degenerate permutations, while configurations such as $|abab\ldots\rangle$ and $|abac\ldots\rangle$, which satisfy the inter- and intra-column constraints and are more numerous, lie higher in energy. However, the ground-state character qualitatively changes in the quantum nematic phase. Fig.~\ref{fig:nematic}d gives the weights of the configurations in the quantum state. 
The classically lowest $|abc\ldots\rangle$ configurations are strongly disfavored, with the state mainly composed of $|abab\ldots\rangle$ configurations, which allow for greater itinerancy between different column states and thus energy lowering via $\hat{\sigma}_x$. 
Note that distant long-range interactions are essential in this analysis, as with truncated interactions all the classical configurations become degenerate.

We can construct an effective one-dimensional spin-1 model of the low-energy sector, and in the
thermodynamic limit the model ground-state generically reproduces the entanglement 
structure seen in Fig.~\ref{fig:nematic}c (see~\cite{supp}). The emergence of
one-dimensional entangled order 
due to kinetic energy and interaction competition recalls other famous nematic phases, for 
example in 2D
fermionic models~\cite{zheng2017stripe}.

\section{Finite phase diagram}
\label{sec:open}

\begin{figure*}
    \centering
    %\includegraphics[width=\textwidth]{figs/finite_phases_12_2.png}
    %\includesvg[width=\textwidth]{figs/finite_phases_12_2.svg}
    \includegraphics[width=\textwidth]{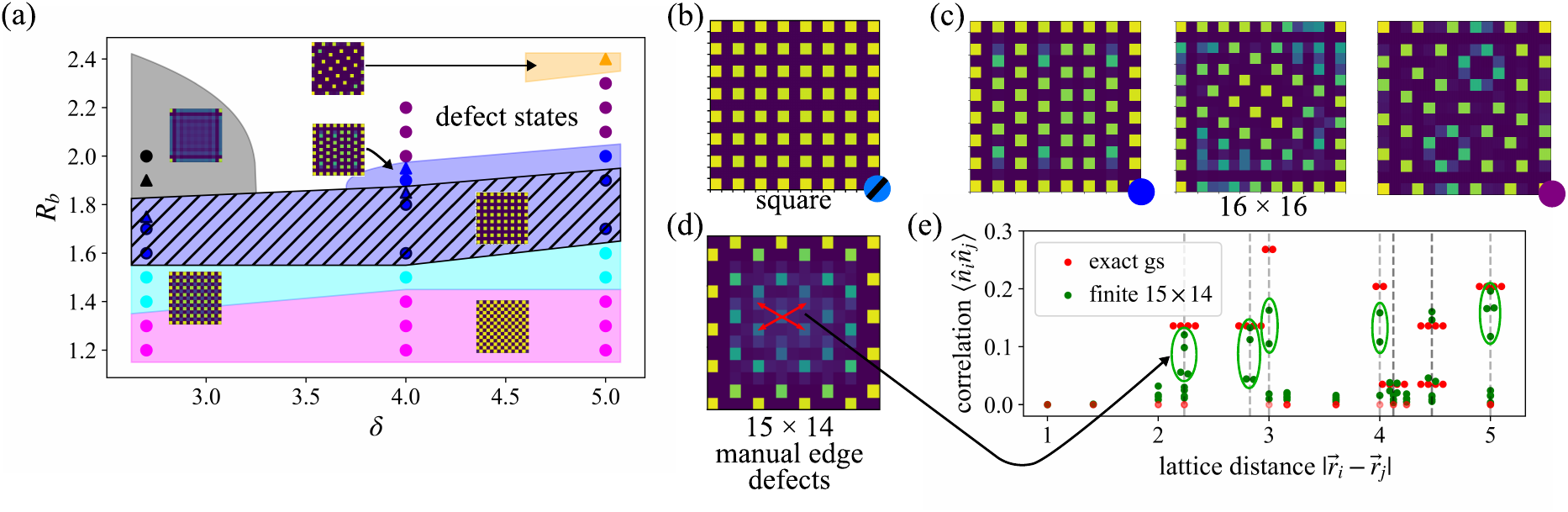}
    \caption{\textbf{Phase diagram of the $15\times15$ finite system and finite lattice orders.}
    (a) The phase diagram, where colors correspond to the same phase classifications as 
    Fig.~\ref{fig:pbc}. Triangles represent tentative classification of points showing inconsistent PEPS convergence, see ~\cite{supp}. A new ``square'' order is specified in (b) and various examples of
    boundary-bulk frustrated ground states in (c).
    (d) The density profile for a nematic-like ground state that can be stabilized on
    a $15 \times 14$ lattice at $(\delta, R_b) = (3.4, 2.1)$ with manually 
    tailored edge excitations (see text).
    (e) Comparing the correlations of the finite nematic phase to the ``exact'' bulk phase. 
    The degeneracy
    of the peaks is split by the boundary excitations, but the number of peaks is generally 
    conserved between the two (green ovals), which provides a clear distinction from mean-field
    states (see Fig.~\ref{fig:nematic}b).
    }
    \label{fig:finite}
\end{figure*}

\begin{figure}
    \centering
    %\includegraphics[width=\linewidth]{figs/expt_compare_final_12_2.png}
    %\includesvg[width=\linewidth]{figs/expt_compare_final_12_2.svg}
    \includegraphics[width=\linewidth]{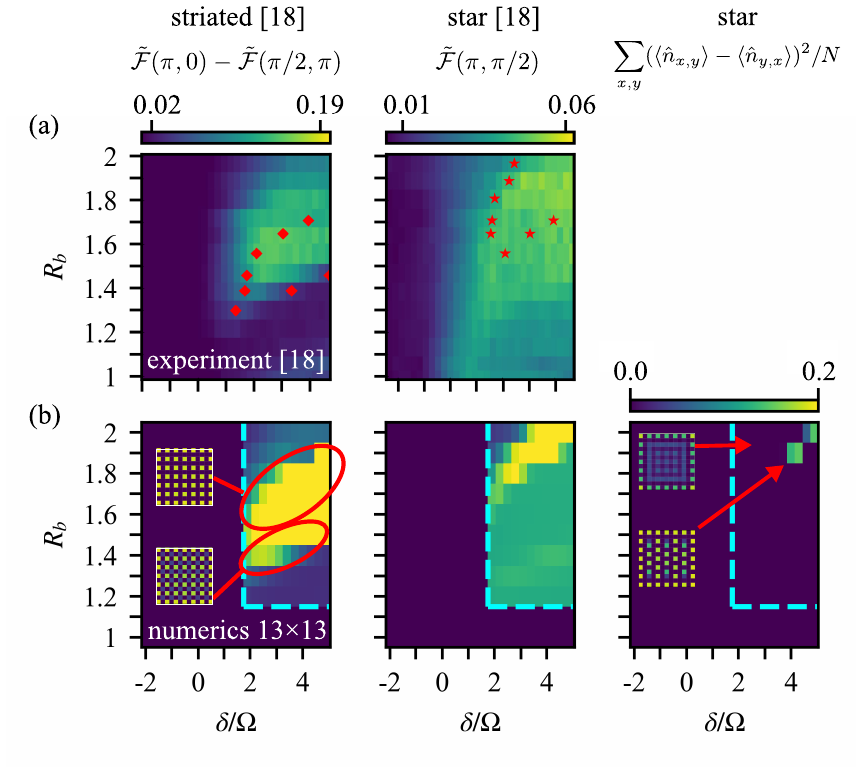}
    \caption{\textbf{Comparison to experiment.} The (a) row directly reproduces experimental phase
    diagram data on the $13 \times 13$ lattice (data extracted from Ref.~\cite{ebadi2020} Fig. 4), 
    while the (b) row 
    is $13 \times 13$ numerical data computed in this work. The first two columns show the order
    parameters used in~\cite{ebadi2020} to identify the square/striated and star phases, while the third
    column shows a new, more sensitive order parameter for the star phase. Red dots in (a) denote
    the phase boundaries assigned in~\cite{ebadi2020}, while the cyan dotted lines in (b)
    indicate the subset of parameter space that was computed. Our calculations support a re-interpretation of the experimental data
    with a significantly larger square/striated region and much smaller star phase.
    }
    \label{fig:expt_compare}
\end{figure}

Current experiments are limited to lattices with open boundary conditions
consisting of a few hundred atoms~\cite{scholl2020,ebadi2020}. To investigate how this modifies the bulk behavior, we computed the phase diagram 
of selected finite lattices from size $9 \times 9$ to $16 \times 16$, using DMRG for the smaller sizes 
and our PEPS methodology for the larger ones.

We first focus (in Fig.~\ref{fig:finite}a) on understanding the 
fate of the ordered phases on the $15 \times 15$ lattice
along three slices: $\delta = 2.7$, 4.0, and 5.0
($16\times 16$ lattice phases, as well as other lattice sizes, are discussed in~\cite{supp}). 
Here, many finite lattice ground state orders resemble those in the bulk. 
However, their regions of stability are substantially reduced and their patterns are broken by frustration. 
Out of the density-ordered quantum phases,
the striated mean-field phase remains due
to its commensurate boundary-bulk configurations, while
in the region of strongest interactions
the nematic phase is destabilized. 
A new region of classical order, called here the square phase (Fig.~\ref{fig:finite}b), 
emerges across much of the $R_b = 1.5 - 1.8$ region where the
star phase was stable in the bulk~\cite{felser2021}. 
We distinguish the square order from the striated order in the
sense that the former has negligible quantum fluctuations on the $(1,1)$-sublattice, 
although it is unclear if the square and striated orders constitute truly distinct phases 
(in the bulk phases the square order is not stable, only the striated order appears).

In Fig.~\ref{fig:expt_compare}, we directly compare the experimental results on the $13 \times 13$ lattice to
our calculations on the same lattice. The analysis of the experiments in Ref.~\cite{ebadi2020} was based on simulations on the $9\times 9$ lattice using truncated interactions. This assigned only part of the experimental non-zero order parameter space to a square/striated phase (see Fig.~\ref{fig:expt_compare}a left panel, note, the order parameter does not distinguish between  square/striated orders). However, our simulations (Fig.~\ref{fig:expt_compare}b left panel) in fact reproduce the full region of the non-zero order parameter, and thus the whole region seen experimentally should be assigned to a square/striated phase, with the square order appearing in the upper part of the region. Similarly, the experimental analysis identified a large region of star order (Fig.~\ref{fig:expt_compare}a middle panel). This assignment is complicated by edge effects, which mean that the order parameter used does not cleanly distinguish the star phase from other phases. However, our simulations suggest that the region of the star phase should be considered to be much smaller, located at the very top of the non-zero order region, 
 and this is confirmed using a different, more sensitive order parameter (Fig.~\ref{fig:expt_compare}b, right panel). Overall, the measured data corresponds more closely to our numerics than earlier simulations, giving confidence in our more precise interpretation (more discussion in~\cite{supp}).

\noindent{\textbf{Stabilizing entangled ground-state order}.} Generally, the impact of boundary physics can be understood in terms of  frustration of the bulk order by the boundary order, where excitations concentrate more densely due to the lower energetic penalty from fewer long-range interactions on the edge.
Examples of the effects of this frustration, ranging from modified bulk orders, to defect dominated states, are shown in Fig.~\ref{fig:finite}b-c (see also~\cite{supp}).

We searched for conditions to stabilize the entangled nematic ground-state on a finite lattice by manipulating boundary effects.
We scanned various rectangular
sizes and explicitly ``removed'' patterns of atoms from the edges
to induce different bulk orders.
We found the best conditions to stabilize the nematic phase occur near
$(\delta, R_b) = (3.4, 2.1)$, on a $15 \times 14$ lattice, while removing edge atoms to create a spacing of 4 on two edges and 3 on the other two edges (Fig.~\ref{fig:finite}d)~\footnote{Note that
the location of this state in phase space 
cannot be directly compared to the locations of 
states in Fig~\ref{fig:finite}a due to the significant difference in the treatment
of the boundary.}.
Although there are strong finite size effects, the density profile  
and correlation functions (Fig.~\ref{fig:finite}d-e) reveal
qualitative similarities to the bulk nematic phase, in particular, the presence of 4-fold correlation peaks at distance $\sqrt{5}$
and $\sqrt{8}$, which are also a feature of the bulk entangled phase (Fig.~\ref{fig:nematic}b).

\section{Conclusion}
Using new tensor network simulation methods, we have obtained a converged understanding of the phase diagram of Rydberg atom arrays in both bulk and  finite simple square lattices. Surprisingly, our bulk phase diagram is quite different from that predicted in earlier numerical studies,
while on finite lattices, our results support a reinterpretation of previous experimental analysis. Theoretically, this is due to the subtle effects of the long-range interactions that are addressed by our techniques, while experimentally, it brings into focus 
the challenge of more accurate theoretical models to interpret increasing experimental capabilities in quantum many-body physics. 
Perhaps most intriguingly, we find that the unfrustrated square lattice supports an entangled quantum nematic phase, brought about the competition between emergent itinerancy and the constraints of the Rydberg interaction. 

A primary focus of Rydberg atom array experiments has been to realize  well-studied short-range Hamiltonians, for example, on frustrated lattices. However, we find that lattice frustration is not necessary to produce interesting entanglement in Rydberg systems. In fact our work highlights the richness and complexity intrinsic to Rydberg atom arrays, due to the non-trivial effects of their native interactions.

\section{Acknowledgements}
M.J.O. acknowledges financial support from a US National Science Foundation Graduate Research Fellowship via grant DEG-1745301. G.K.C. acknowledges support from the US National Science Foundation via grant no. 2102505. Computations were conducted in the Resnick High
Performance Computing Center, supported by the Resnick Sustainability 
Institute at the California Institute of Technology. DMRG calculations
were performed with the ITensor library~\cite{itensor}. PEPS calculations
were performed using quimb~\cite{quimb} with PyTorch as the
backend~\cite{pytorch}.

%\detailtexcount{main}

\nocite{Mclean2018}
\nocite{nocedal2006numerical}
\nocite{vidal2003efficient}
\nocite{jiang2008accurate}
\nocite{corboz2010simulation}
\nocite{mcculloch2008infinite}

% include the below on local machine, but comment out for submission
%\bibliographystyle{apsrev4-1}
%\bibliography{refs}

% include the below for submission, make sure above is commented out

\clearpage
%\newpage
%\mbox{}
%\clearpage
%\documentclass[reprint,aps,twocolumn,groupedaddress]{revtex4-1}
%\usepackage{color}
%\usepackage{graphicx}
%\usepackage{enumerate}
%\usepackage{amsfonts,amssymb}
%\usepackage{amsmath}
%\usepackage{braket}
%\usepackage{verbatim}
%\usepackage{bbold}
%\usepackage{svg}

%\newcommand{\BLUE}[1]{{\color{blue} #1}}
%\newcommand{\RED}[1]{{\color{red} #1}}
%\newcommand{\dens}[0]{\langle \hat{n} \rangle}

%\begin{document}

\title{Supplementary Information: 
Entanglement in the quantum phases of an unfrustrated Rydberg atom array}
%\author{Matthew J. O'Rourke}
%\affiliation{\footnotesize{Division of Chemistry and Chemical Engineering,
%California Institute of Technology, Pasadena, CA 91125, USA}}
%\author{Garnet Kin-Lic Chan}
%\affiliation{\footnotesize{Division of Chemistry and Chemical Engineering,
%California Institute of Technology, Pasadena, CA 91125, USA}}
\bigskip
\date{\today}

\maketitle

\section{Numerical methods}

This section gives details for the
numerical simulations in this work. Principally, it
will focus on algorithmic subtleties and sources of error, as well as 
the strategies employed to resolve the physics of the Rydberg atom
system.

\subsection{$\Gamma$-point DMRG}

\begin{figure*}
    \centering
    %\includesvg[width=0.95\linewidth]{figs/supp/pill_phase_convergence.svg}
    %\includegraphics[width=0.95\linewidth]{figs/supp/pill_phase_convergence.png}
    \includegraphics[width=0.95\linewidth]{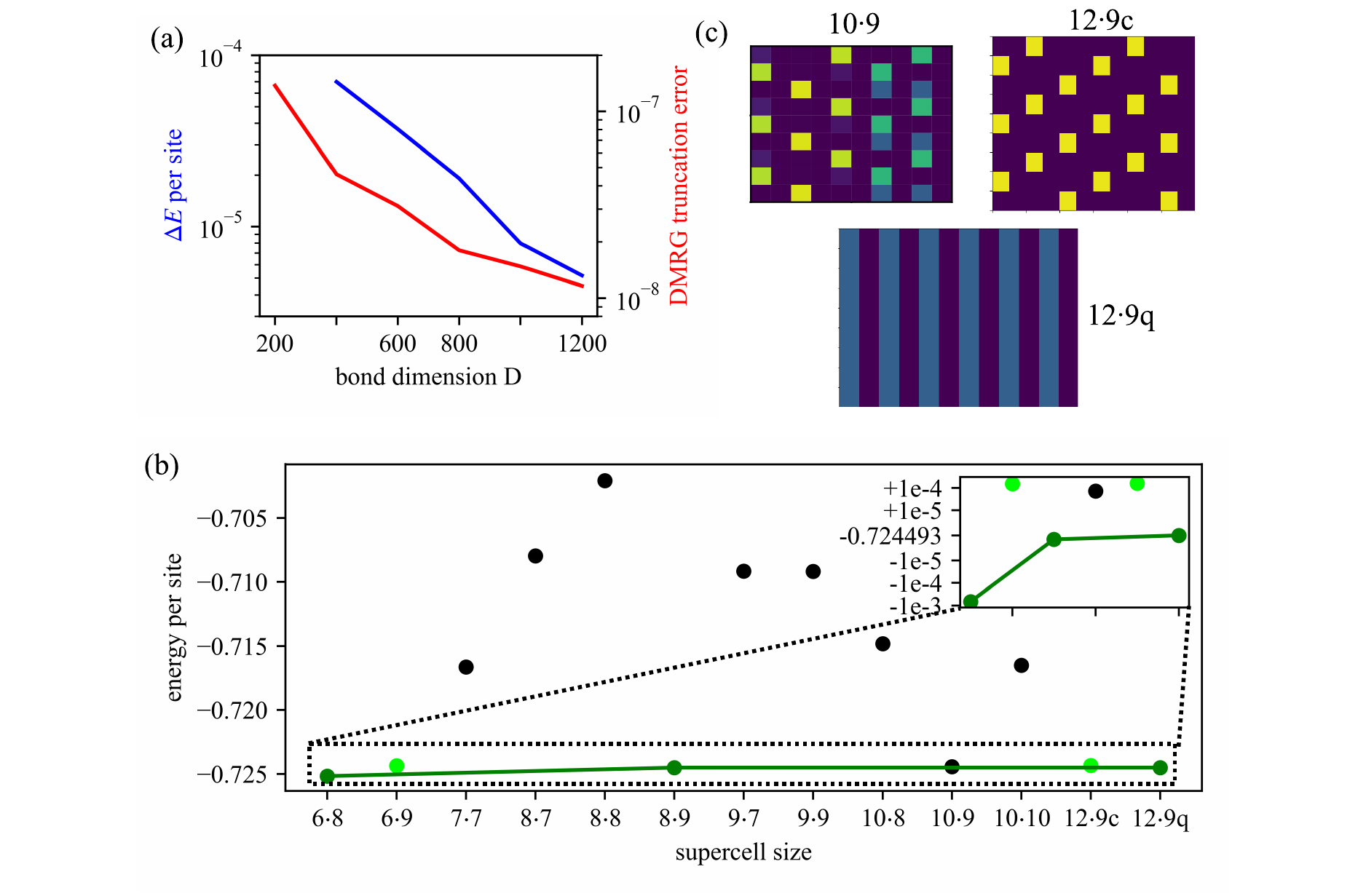}
    \caption{Convergence of $\Gamma$-point DMRG in the most difficult region
    of the phase diagram $(\delta, R_b) = (5.0-6.0, 2.3)$. (a) Shows the convergence w.r.t.
    bond dimension of the largest truncated DMRG singular value (red) and the change 
    in energy per site relative to the energy obtained with bond dimension $D-200$ (blue). 
    (b) The energies per site of a large variety of supercell sizes. This is adapted from
    Fig. 1 of the main text to highlight the relevant points.
    The connected dark green
    points are the nematic phase, and lime green points are the low energy 3-star $\frac{1}{6}$-density
    crystalline phase. The inset shows the convergence of the nematic phase energy w.r.t. 
    supercell size and gaps to the other low energy solutions, whose density profiles are
    shown in (c). Note that, between (a) and (b),
    the nematic phase is converged to below $10^{-5}$ accuracy while the competing states
    differ in energy by at least $10^{-4}$.}
    \label{fig:gamma_pill_conv}
\end{figure*}

\begin{figure}
    \centering
    \includegraphics[width=0.95\linewidth]{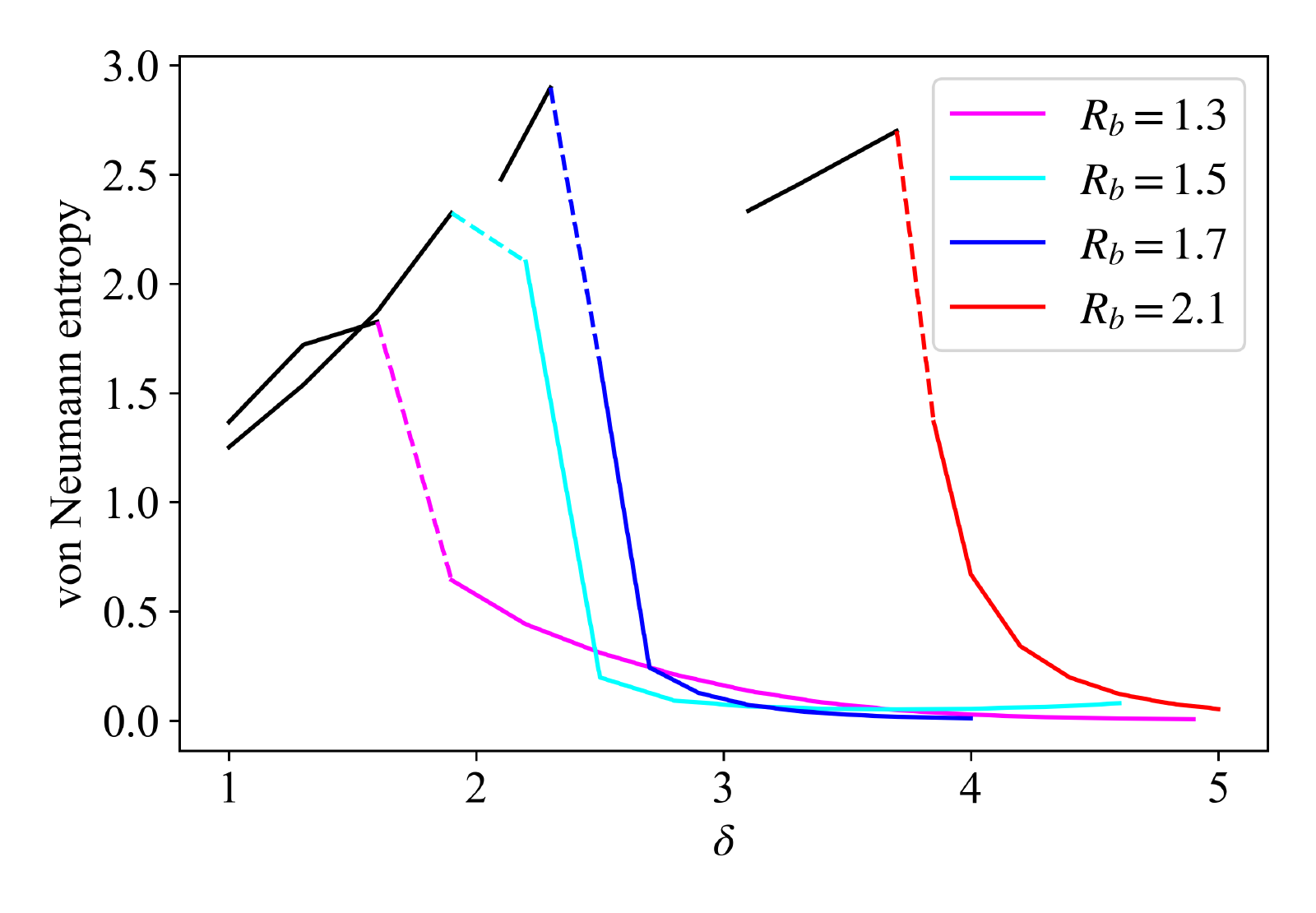}
    \caption{Bipartite entanglement entropy of various crystalline phases as $\delta$
    increases. Each line is a slice over $\delta$ values for a fixed $R_b$ value.
    Black line segments denote when the ground state is in the disordered
    phase. Solid colored line segments denote when the ground state is an ordered
    crystalline phase (same color classifications as the phase diagram in main text).
    Dotted line segments denote the ``transition zone'' of a given line
    between the disordered phase and an ordered phase. These are simply a result of
    the finite resolution used to sample phase space in the phase diagrams.}
    \label{fig:crystalEE}
\end{figure}

\subsubsection{Convergence and physical strategy}
The $\Gamma$-point DMRG uses a lattice sum construction of the interaction
terms in order to approximate the bulk physics of the Rydberg system (see Eq. 2, main text). 
As explained in the main text, this is quite different from 
the
cylindrical boundary conditions
employed in previous studies~\cite{lukin2020,verresen2020prediction,samajdar2021quantum}
as the represented system is truly infinite. In particular, to increase the range of interactions, we do not need to incur the exponential increase in cost that arises from the associated increase in cylinder width in the standard cylindrical approach.

We find that we can converge our calculations to sufficiently high accuracy
with reasonable bond dimensions.
Even in the very complicated region of the phase diagram near $\delta=5.0-6.0, R_b=2.3$,
we can distinguish the ground-state orders using a bond dimension of $D=1200$, 
as shown in Fig.~\ref{fig:gamma_pill_conv}. However, although this is enough to identify the ground state order, higher bond dimensions would be needed to capture the phase transitions with high precision; given the large region of phase space explored here, we leave such detailed calculations to future work.

The strategy used to generate the bulk phase diagram in main text Fig. 2a, as well 
as the truncated interaction phase diagram Fig. 2b, is as follows. 
\begin{itemize}
\item For a given point
in phase space $(\delta, R_b)$, run a $D_{\mathrm{max}}=1000$ simulation for all reasonable
supercell sizes between $4 \times 4$ and $10 \times 10$, as well as $12 \times 9$.
\item Identify
all supercells for which the ground state has an energy per site within $10^{-2}$ 
of the lowest energy. 
\item If there are competing orders, ensure these solutions are 
all sufficiently converged by requiring (i) the largest singular value truncated during 
the final DMRG sweeps is less than $10^{-8}$, 
and (ii) corrections to the energy when increasing supercell size (up to $12 \times 9$ maximally) 
are smaller than the energy gap between competing states (Fig.~\ref{fig:gamma_pill_conv}).
\item The ground state phase is then identified by evaluating simple density-based 
order parameters on the largest supercell size which hosts the ground state order.
\end{itemize}
The only time this convergence criteria is not satisfied is for disordered phase solutions
near the order-disorder phase transition (largest truncated DMRG singular value is $\sim 10^{-6}$), 
for which all large supercells show a disordered solution. The classification of the phase
in this region is supplemented by analyzing the ground state entanglement entropy,
which shows a distinctive ``drop'' when the phase becomes ordered (see Fig.~\ref{fig:crystalEE}).

Importantly, this strategy completely neglects possible orders with unit cells larger than $10 \times 10$
or $12 \times 9$, as well as non-periodic solutions. 
Although orders with unit cells of this large size are not expected in the 
region of the phase diagram under investigation in this work due to the relatively
high crystal densities (and thus close spatial packing)~\cite{lukin2020,lauchli2019},
our study cannot definitively rule out the stability of such solutions.

\subsubsection{Finite size errors}
There are a two main sources of finite size error in this formulation of the bulk system.
The first comes from the lattice sum form of the long-range interaction. Given the $\Gamma$-point Hamiltonian in Eq. 2 (main text), there
are some interaction terms of the form $\hat{n}_i \hat{n}_i / |\vec{r}_i - \vec{r}_{i+R_l}|$.
These represent the interaction of a Rydberg excitation with its own periodic ``image'' in a distant
supercell. For a classical crystal, this image term is exact, but
in a quantum phase, it is an approximation. Due to the idempotency of $\hat{n}$, this term simplifies to
$\hat{n}_i / |\vec{r}_i - \vec{r}_{i+R_l}|$, which would not exist in the Hamiltonian
if the supercell were large enough to contain both points $i$ and $i+R_l$. The effect of
this error on the energy \textit{per site} can be estimated by the quantity,
\begin{equation}
    \Delta e = \frac{2 \cdot R_b^6}{\rho_{ex} \cdot \min(L_x, L_y)^6} (\langle \hat{n}_i
    \rangle - \langle \hat{n}_i \rangle^2).
\end{equation}
Here, $\langle \hat{n}_i \rangle$ is the expectation of the local Rydberg excitation for
a single characteristic excited site, while $\rho_{ex}$ is the density of sites which have
the characteristic excitation of $\langle \hat{n}_i \rangle$.
Note that $\Delta e$ is always positive, can be 
systematically reduced 
by increasing the supercell size, and it is always close to 
0 for (almost) classical crystals with excitation densities close to 1 or 0, regardless of cell size. 

The other source of systematic error comes from the constraint
on the wavefunction imposed by
approximating a bulk system by a supercell. Most obviously, this means
that certain orders cannot appear in smaller supercell, even for classical crystals. In the case of quantum orders, even for a fixed order there are finite size effects on the emergent kinetic energy of defects.

\subsection{PEPS}
\label{sec:peps_method}
\begin{figure}
    \centering
    \includegraphics[width=0.95\linewidth]{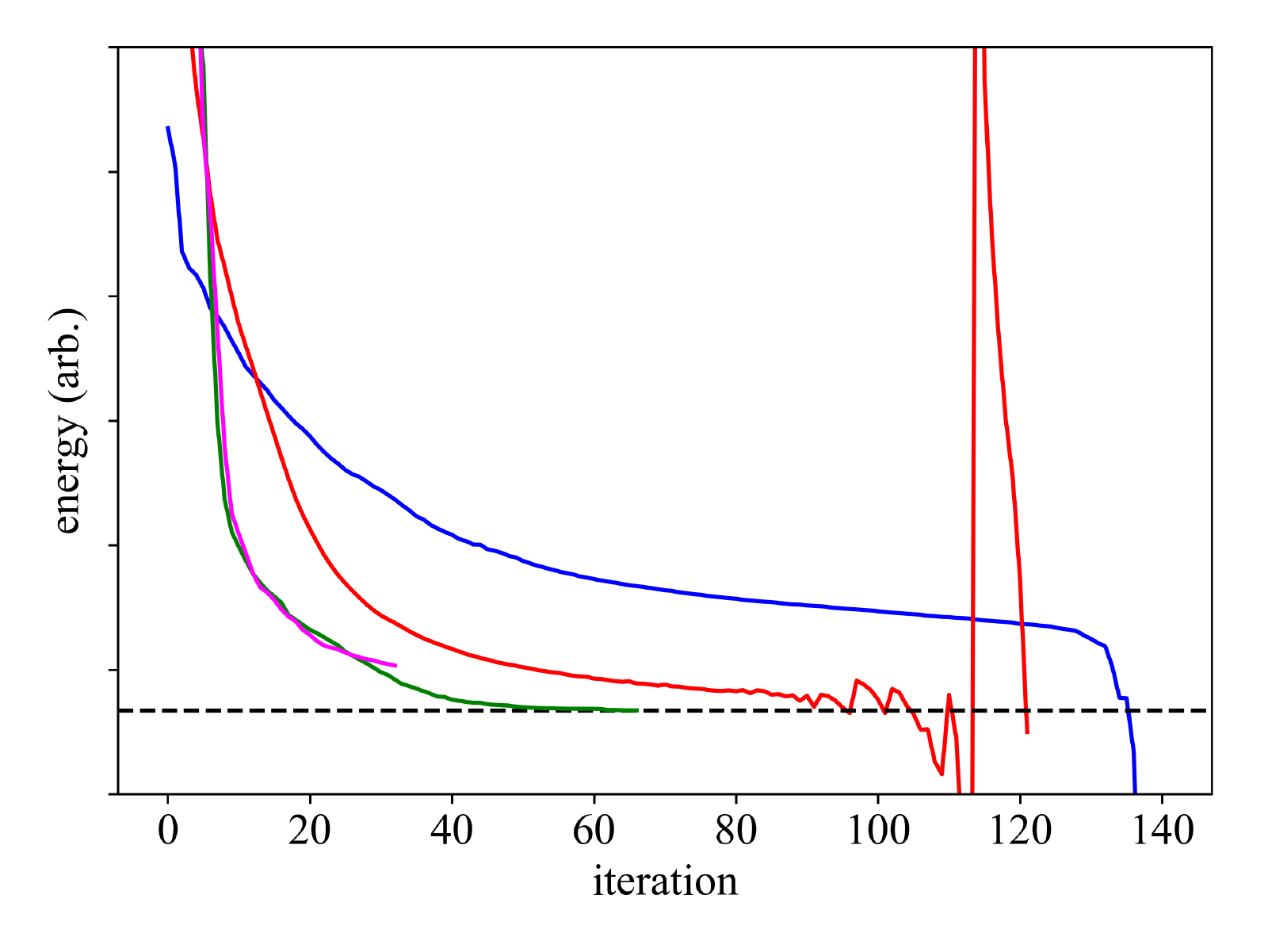}
    \caption{Examples of typical optimization trajectories for long-range
    PEPS using different automatic differentiation schemes. 
    The blue line often occurs with a 
    naive implementation of the energy evaluation algorithms and use of
    a line search which does not minimize gradient norm. The red line can occur even when using a more sophisticated energy evaluation including local norms and/or
    a multi-evaluation cost function. The stable magenta and green lines result from combining the four techniques discussed in Sec.~\ref{sec:peps_opt}.
    The difference between the magenta and green curves reflects the quality of the initial guess.}
    \label{fig:peps_traj}
\end{figure}

The PEPS simulations in this work
combine recent advances in
optimizing PEPS wavefunctions using automatic differentiation~\cite{autodiffTN} and
2D operator representations of long-range interactions~\cite{gmpos}.  This combination illuminated many new challenges for PEPS optimization with respect to complicated Hamiltonians.
This section will detail the various challenges and the technical solutions used in this work. The instability of PEPS optimization remains an open problem and it is an area of future research to determine a PEPS optimization pipeline (using automatic differentiation) that is fully robust to problem instance. In this section, $D$ will refer to the PEPS
bond dimension and $\chi$ will refer to the maximum bond dimension allowed during contraction
before approximations (via SVD) are performed.

\subsubsection{Operator representation}
The method proposed in Ref.~\cite{gmpos} to represent Hamiltonians with long-range interactions 
writes the interaction potential as a sum of Gaussians,
\begin{equation}
    \frac{1}{(\sqrt{x^2 + y^2})^6} \approx \sum_{k=1}^K c_k e^{-\lambda_k (x^2 + y^2)}
    \equiv V_{\mathrm{fit}}(\vec{r}).
\end{equation}
Using the methods
in Ref.~\cite{Mclean2018}, we can obtain a $K=7$ fit with error 
$\epsilon = \max_i|1/\vec{r}_i^6  - V_{\mathrm{fit}}(\vec{r}_i)| = 10^{-5}$ on the 
domain $\vec{r} \in [1, 16 \sqrt{2}]$, which is used throughout the work.

\subsubsection{Essential computational techniques}
As originally discussed in Ref.~\cite{autodiffTN}, when trying to use automatic differentiation to optimize a PEPS there are a few essential techniques that must
be employed, which are not typically ``default'' in standard automatic differentiation
libraries. They are ``essential'' in the sense that without them the computation of the energy 
expectation value and its derivative will typically not run to completion due to
out-of-memory errors or numerical infinities. 
These techniques are:
\begin{itemize}
\item Numerical stabilization of the gradient of SVD, by adding Lorentzian broadening to the inverse
singular values.
\item Significant usage of ``checkpointing'' when evaluating the energy to reduce the
memory load of computing gradients.
\end{itemize}
Both of these techniques are explained in significant detail in Ref.~\cite{autodiffTN}.

\subsubsection{Stabilizing the optimization}
\label{sec:peps_opt}
A straightforward implementation of the energy expectation value as described in~\cite{gmpos},
with optimization via automatic differentiation including the above techniques, typically
fails to find the ground state PEPS for the Rydberg Hamiltonian (see Fig.~\ref{fig:peps_traj}). 
This failure can be generally attributed to the fact that in the quantity under optimization
$E = \frac{\langle \psi | H | \psi \rangle}{\langle \psi | \psi \rangle}$, both the numerator
and denominator are evaluated approximately and thus the computation is not strictly 
bound by the variational principle. 
Consequently, the optimization can find pathological regions of the 
PEPS parameter values which make the PEPS contractions
inaccurate for the chosen $\chi$, even when starting from an accurately contractible PEPS.
Unfortunately, in this problem we find that simply raising the value of $\chi$ does not prevent this behavior
until $\chi$ is impractically large.

In order to mitigate this problem 
we use the following four techniques in tandem:
\begin{itemize}
\item We employ line search methods that minimize the gradient norm 
as well as the energy. In this work, we use the BFGS algorithm~\cite{nocedal2006numerical} in conjunction with such a line search, as suggested in~\cite{autodiffTN}.

\item We use the cost function  $E_1 / 2 + E_2 / 2 + \lambda |E_2 - E_1|$
where $E_1$ and $E_2$ are the energies of PEPS on lattices rotated
by 180 degrees and $\lambda$ is a penalty factor. This strongly
penalizes the optimization from entering parameter
space with large contraction error (where $E_1$ and $E_2$ would be very different).

\item During the first iterations of the gradient optimization we only update small patches of tensors at a time, which are chosen to break spatial symmetries that may be contained in the
initial guess. After this has pushed the optimization towards the symmetries of the true
ground state order, then all tensors can be updated at each optimization step.

\item We evaluate the numerator and denominator of $E$ in a consistent way by using 
``local normalization'' during the computation of $\langle \psi | H | \psi \rangle$.
This means that, writing $H$ as a comb tensor sum $H = \sum_{i=1}^{L_x} h_i$, then for each comb
tensor numerator $\langle \psi | h_i |\psi\rangle$, the associated denominator uses the identical
contraction, but with $h_i$ replaced by the identity (the environments are not recomputed).

\end{itemize}

Combining all four of these techniques removes the most egregious instabilities in the optimization trajectory (see Fig.~\ref{fig:peps_traj}), at the cost of a slightly larger 
computational burden. However, as in more standard DMRG calculations with small bond dimension, convergence to the correct ground-state (rather than a local minimum) still requires a reasonable initial guess.

\subsubsection{Initial guess}
Obtaining an accurate ground state PEPS typically relies on starting with an accurate initial guess.
The predominant algorithms to generate such a guess for problems with a local Hamiltonian
are simple update~\cite{vidal2003efficient,jiang2008accurate,corboz2010simulation} 
or imaginary time projection of a converged small $D$ solution to a larger $D$ guess. However,
in the presence of long-range interactions it becomes challenging to generalize either of
these methods in an efficient and/or accurate way. We therefore 
used the following simple
scheme to generate initial guesses in this work.
\begin{itemize}
    \item Sum $n$ manually constructed $D=1$ PEPS to obtain an initial PEPS of bond dimension $D=n$.
    The configurations of these $D=1$ PEPS were 
    set to reproduce specific low energy Rydberg crystals 
    and defects within them.
    
    \item For small $R_b$: truncate the long-range interactions in $H$ to next-nearest, 
    or next-next-nearest, neighbor interactions (distance of $\sqrt{2}$ or 2), and run 
    conventional simple update starting from the above manually summed PEPS. This fails once
    the ground state excitations are spaced by more than 2.
    
    \item For large $R_b$: add positive random noise to the manually summed PEPS, 
    and then run a highly approximate, first-order gradient optimization for $\sim 25$ iterations
    using a large step size when updating the parameters.
\end{itemize}

\subsubsection{Convergence and physical strategy}

\begin{figure*}
    \centering
    %\includesvg[width=0.95\linewidth]{figs/supp/finite_phase_diag_supp.svg}
    \includegraphics[width=0.95\linewidth]{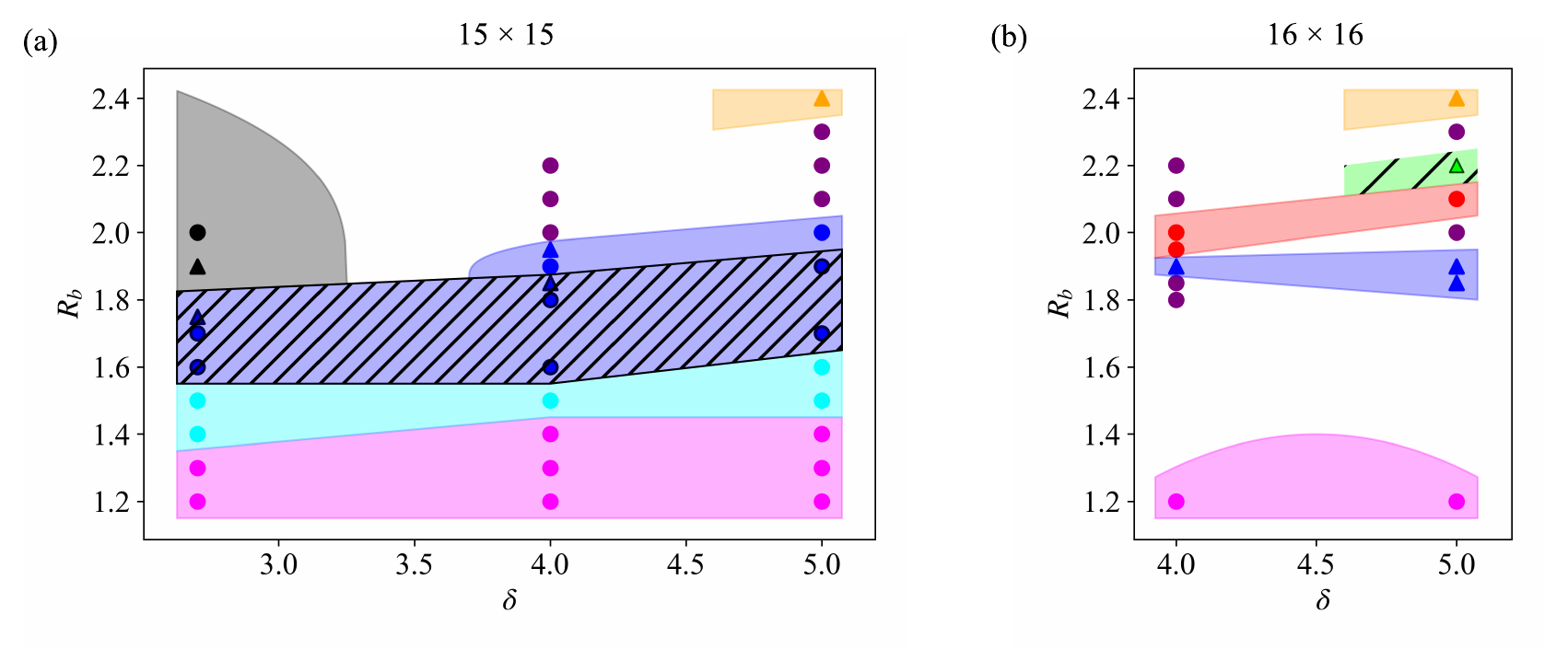}
    \caption{Phase diagrams of $15 \times 15$ (a) and $16 \times 16$ (b) arrays, detailing 
    convergence. Circular points indicate systematic convergence with PEPS up to bond dimension
    $D=5$, while triangles indicate intermittent convergence with PEPS, requiring
    supplemental convergence checks using 2D DMRG. More details are available in
    Sections~\ref{sec:finite_details} and~\ref{sec:peps_method}. 
    Colors used in these plots correspond identically
    to the colors used in the main text to identify phases.
    }
    \label{fig:finite_pdiag}
\end{figure*}

\begin{figure}
    \centering
    \includegraphics[width=\linewidth]{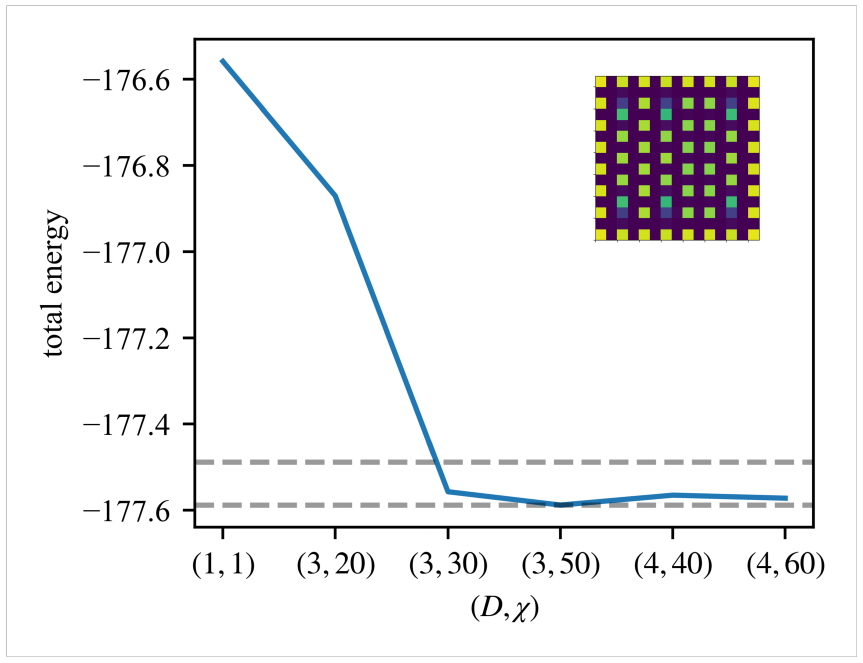}
    \caption{An example of systematic convergence of PEPS on the 
    $15 \times 15$ lattice for the frustrated
    star phase at $\delta=4.0$, $R_b = 1.9$. 
    The region between the horizontal lines indicates a change in energy of $0.01\%$ relative to
    the lowest obtained value. The PEPS is deemed converged because many simulations with
    increasing $(D, \chi)$ return energies that fall within this region.
    Note that the star phase, like most ordered phases, is sufficiently converged by 
    $D=3$ due to the predominant mean-field character of most ordered phases 
    (discussed in main text).
    }
    \label{fig:peps_converge}
\end{figure}

Despite the simple procedure to generate initial guesses,
we were usually able to systematically converge PEPS solutions according to the conventional 
protocol of increasing $D$ and $\chi$ until the energies corresponding
to multiple increasing $(D, \chi)$ pairs all vary by less than $0.01\%$ relative to each other.
(e.g. see Fig.~\ref{fig:peps_converge}).
In this study, we used maximal values of $D=5$, $\chi=100$.
However, for a small number of phase points $(\delta, R_b)$ we encountered inconsistent 
convergence of PEPS solutions (see Fig.~\ref{fig:finite_pdiag}), where increasing $D$ and $\chi$ 
did not systematically result in finding a PEPS with a lower energy,
instead getting stuck in various local minima.
We attribute this to the low quality of the initial guesses for larger $D$ and $R_b$.

In these cases when PEPS \textit{energies} could not be systematically converged to within $0.01\%$,
the observed \textit{order} of the various low-energy solutions were nonetheless the same.
The differing energies arose due small quantitative differences such as single-site defects and
variations in the local density $\langle \hat{n}_i \rangle$.
To further increase certainty in the observed order, we also 
compared the PEPS solutions to the results of 2D DMRG on the same finite lattice, since
the convergence properties of DMRG are much more well-understood. In all cases, the low-energy 
PEPS solutions had similar energies to the approximate DMRG (relative difference $< 1\%$), 
and they all 
showed the same generic low-energy ground state order.
The energy gap between phases appeared to be sufficiently large to allow for a tentative
classification of the
order of this small number of phase points, even though the DMRG was not necessarily 
converged to high precision (due to the wide lattices) and the PEPS convergence 
could not be definitively confirmed. The uncertainty in convergence highlights 
remaining challenges in simulating complex large 2D interacting problems with 
competing phases using tensor network techniques.
The relevant points in the finite lattice phase diagram are labelled by triangles in
Fig.~\ref{fig:finite_pdiag} above, and in Fig. 4 of the main text.

\subsection{Finite 2D DMRG}

\begin{figure}
    \centering
    \includegraphics[width=\linewidth]{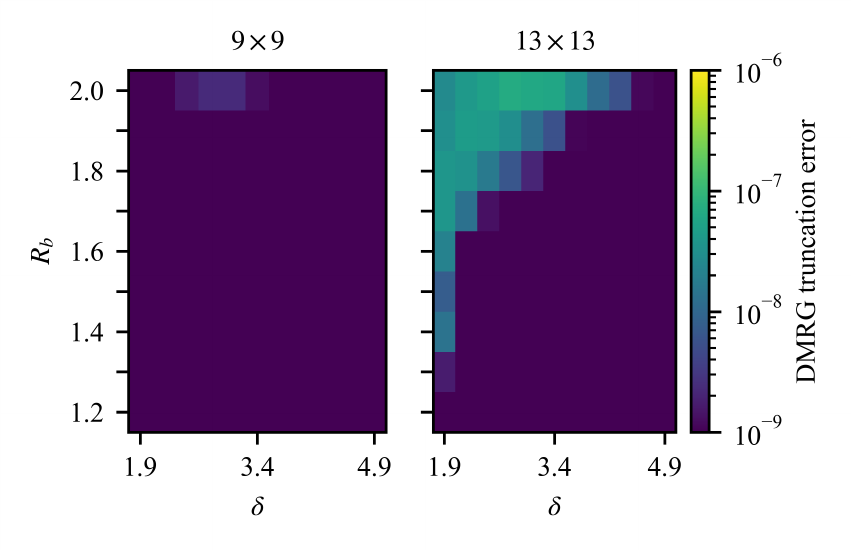}
    \caption{Accuracy of 2D DMRG on the $9 \times 9$ and $13 \times 13$ finite lattices
    (open boundaries). 
    The displayed regions of parameter space correspond exactly
    to the computed regions in Fig. 5 (main text) and Fig.~\ref{fig:expt_compare_supp} (SI).
    The reported error is the largest truncated singular value
    during the final DMRG sweeps (i.e. once converged).
    Note that in the ordered regions the error is $\sim 10^{-9}$, and it grows to $\sim 10^{-7}$ as
    the ground state becomes disordered on the $13 \times 13$ lattice due to increasing 
    entanglement.
    }
    \label{fig:dmrg_err_9x9}
\end{figure}

Standard 2D DMRG calculations with open boundaries were used to study the 
$9 \times 9$ system, a low-entanglement region of the $13 \times 13$ system, and to
supplement convergence of PEPS on the larger $15 \times 14$, $15 \times 15$,
and $16 \times 16$ lattices.
Like the PEPS calculations, these too included
all long-range interactions (according to Eq. 1 in the main text). 
The maximal bond dimension used for the $9 \times 9$ and $13 \times 13$
simulations was $D_{\mathrm{max}} = 1200$, which we found was more than enough
to accurately study the regions of interest in Fig. 5 (main text) for these lattices
(see Fig.~\ref{fig:dmrg_err_9x9}). For supplementing PEPS convergence on the larger lattices,
we used $D_{\mathrm{max}} = 750$. Although this bond dimension is not large enough to
capture the ground state energy or entanglement of such large systems with high precision,
we found it sufficient to capture
the first 3-4 digits of the ground state energy and to help with
distinguishing between the different low-entanglement ordered
phases present in the finite phase diagram, which have substantially larger gaps than the bulk system due to edge effects.

\subsection{Mean field and classical}
The mean field phase diagram for the bulk system (including all long-range
interactions) in Fig. 2d (main text) was generated by the following procedure.
\begin{itemize}
\item Parameterize the single site wavefunction as 
$| \phi_i \rangle = \sin^2(\theta_i) |0 \rangle + \cos^2(\theta_i) |1\rangle$,
where $|0 \rangle$ is the atomic ground state and $|1 \rangle$ is the excited Rydberg
state. 
\item Construct a completely un-entangled many-body wavefunction as a typical product
of these single-site states according to all reasonable unit cells between size 
$2 \times 2$ and $8 \times 10$ (supercells are not necessary for mean-field convergence).
\item Initialize all possibly relevant configurations for each unit cell as initial guesses.
\item Minimize the $\Gamma$-point energy for all guesses with respect to the $\{ \theta_i \}$ using
gradient descent. Analytic gradients are easily derived, or automatic differentiation
can be employed.
\item Classify the phase of the lowest energy state using the same density-based order parameters
as the $\Gamma$-point DMRG calculations. 
\end{itemize}
The phase space was scanned with a $\delta$-resolution 
of 0.1 and a $R_b$-resolution of 0.025. 
Importantly, these calculations are subject to the same
limitation as the $\Gamma$-point DMRG - they do not capture any possible low energy states with
a unit cell larger than $8 \times 10$.  Although such states are not expected in the phase space
under examination, this study cannot definitively rule them out.

The classical phase diagram for the bulk system (including all long-range interactions) in
Fig. 2c (main text) was generated by the following procedure. 
\begin{itemize}
\item Run classical Monte Carlo minimization
of the $\Gamma$-point energy for every unit cell size between 
$2 \times 2$ and $10 \times 10$ at phase space points spaced by $\Delta \delta = 0.3$, 
$\Delta R_b = 0.1$. 
\item For all low energy configurations obtained at all phase points,
derive their continuous functional form $E(\delta, R_b)$ by numerically integrating
the interactions. 
\item Analytically solve for the
intersection line between each adjacent pair of configurations in phase space
that have minimal energy. 
\end{itemize}
These calculations are also subject to the same limitation as 
above - any states with unit cells larger than $10 \times 10$ are not captured, and
we cannot rule out their possible existence.

\section{Bulk phase diagram degenerate region}
\begin{figure}
    \centering
    \includegraphics[width=0.95\linewidth]{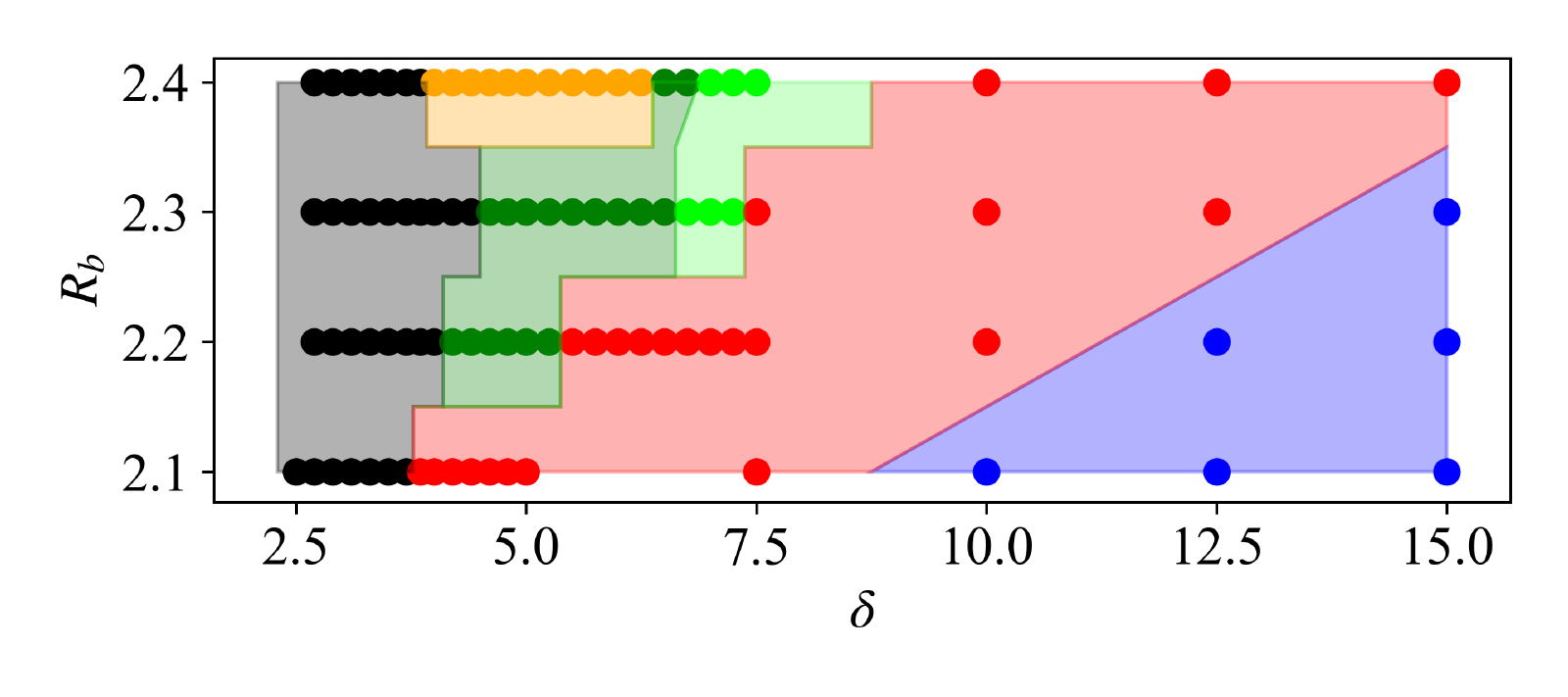}
    \caption{Expanded view of the large-$R_b$ part of the bulk phase diagram,
    computed with $\Gamma$-point DMRG.
    For $\delta \leq 5.0$, this data is identical to Fig. 2a in the main text. All
    colors correspond to the same phases as in the main text Fig. 2. The small
    lime green region indicates the degenerate zone where the gap between the
    3-star and nematic phases becomes very small.}
    \label{fig:pbc_top}
\end{figure}

In the main text it was briefly mentioned that there is a small region of the bulk phase diagram
where the nematic phase and 3-star phase become essentially degenerate. By this we mean that
their gap becomes too small to resolve within the estimated finite size error in the
$\Gamma$-point DMRG numerics. Using the $\Delta e$ finite size error measure defined above, 
 for $[R_b = 2.3, L_x=12, L_y=9]$, we have $\Delta e \approx 3 \cdot 10^{-5}$ in the
nematic phase and $\Delta e \lessapprox 8 \cdot 10^{-6}$ in the 3-star phase.
An expanded view (in $\delta$) of the upper part of the bulk phase diagram is shown in 
Fig.~\ref{fig:pbc_top}. The degenerate region emerges between the nematic
phase and the $\frac{1}{5}$-staggered phase near $\delta=7.0$, as indicated by the lime green color.

\section{Bulk phase transitions}
\begin{figure}
    \centering
    \begin{tabular}{c}
        \includegraphics[width=0.75\linewidth]{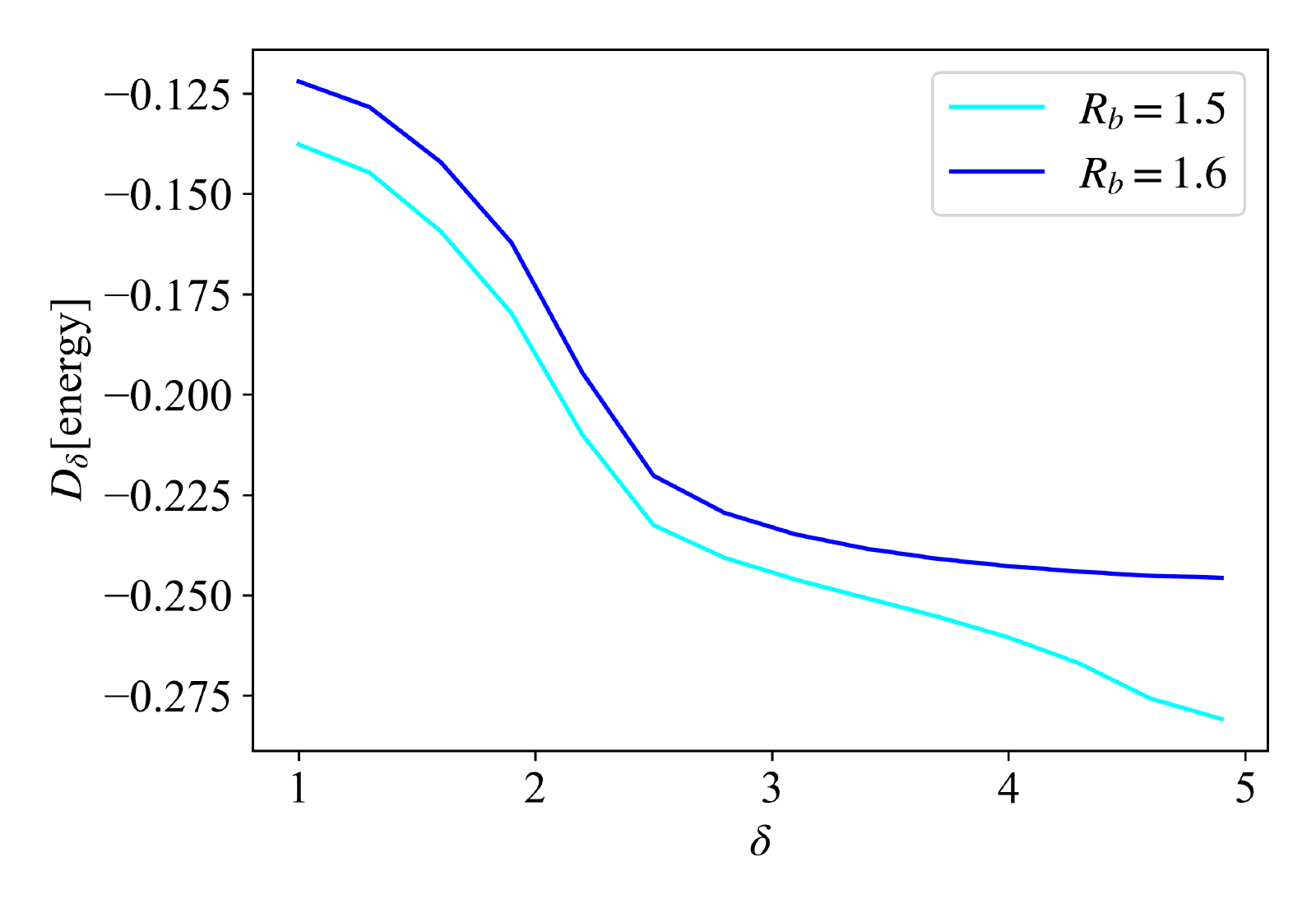} \\
        \includegraphics[width=0.75\linewidth]{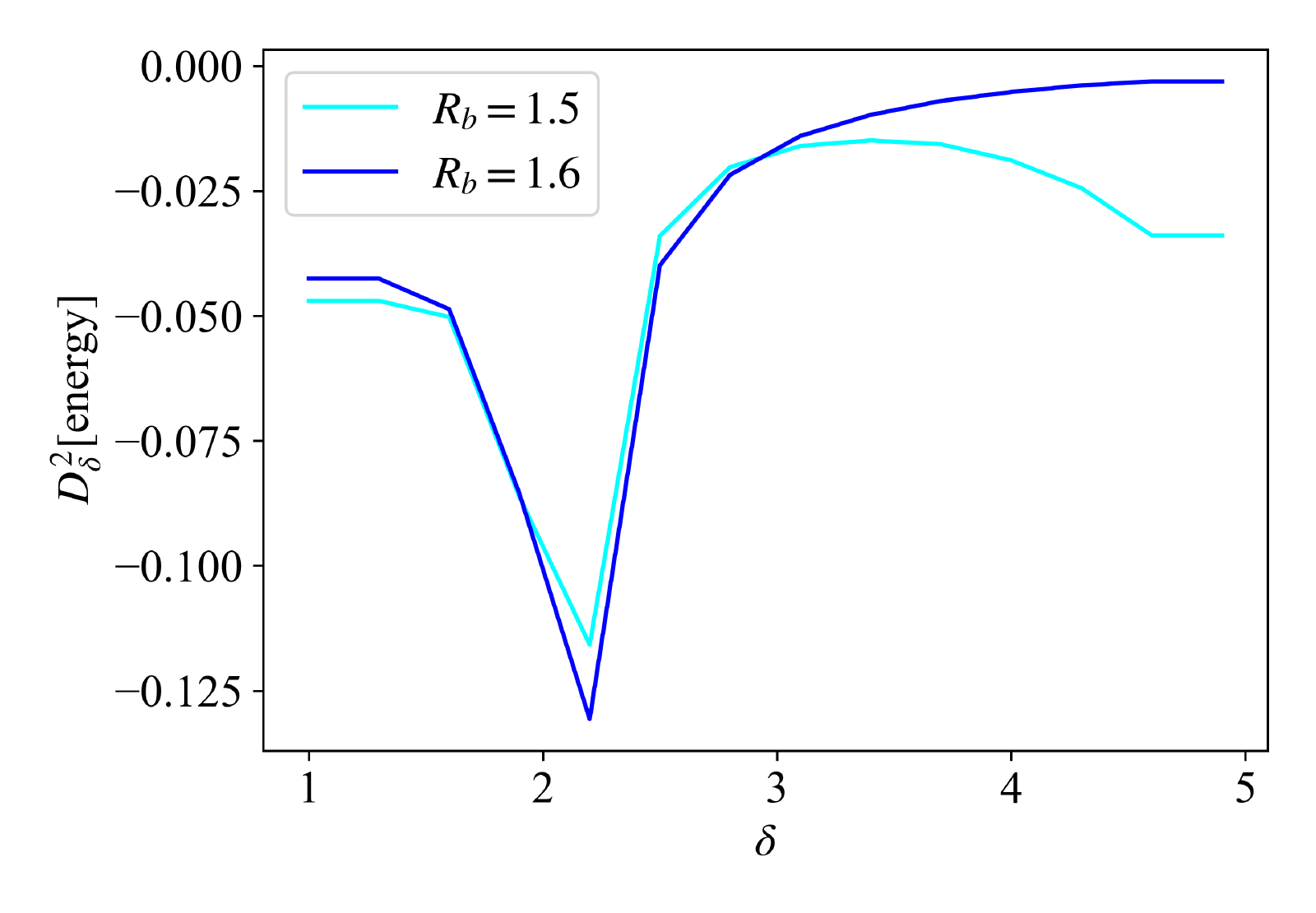}
    \end{tabular}
    \caption{Numerical evidence of second-order phase transitions between
    the disordered phase and the star (blue) and striated (cyan) phases. 
    \textit{Top:} First derivative of the energy with respect to $\delta$. 
    \textit{Bottom:} Second derivative of the energy with respect to
    $\delta$. Both are estimated using standard finite difference formulas.}
    \label{fig:transition_order}
\end{figure}

The order-disorder phase transitions that occur throughout the bulk phase diagram have
been characterized as continuous phase transitions in previous work~\cite{lukin2020}.
Although full, precise characterization of all bulk phase transitions is beyond the 
scope of this work, we are able to estimate the order of some transitions using 
straightforward numerical differentiation of the energies. Figure~\ref{fig:transition_order}
shows the first and second derivatives of the energy as a function of $\delta$, for various
values of $R_b$. The clear peaks in the second derivatives near the critical values of $\delta$
support previous conclusions that the disorder$\rightarrow$star and 
disorder$\rightarrow$striated phase transitions are indeed second-order.

\section{1D model of the bulk nematic phase}
\label{sec:1dmodel}
In the main text the character of the nematic phase was discussed in terms of
the classical configurations that make up the quantum wavefunction. It was pointed out
that all the low-energy (and thus the most relevant) classical configurations can
be described in a succinct notation like $\ket{abcabc...}$ in terms of
compositions of 3 individual column states
$\ket{a}, \ket{b}$, and $\ket{c}$ which are defined in Fig. 3 of the main text.

This notation is very suggestive of the idea that a nice model for the 2D state can
be written as a 1D MPS with a local Hilbert space of dimension 3, spanning 
$\ket{a}, \ket{b}$, and $\ket{c}$. This model ignores the microscopic details of
how one column state $\ket{a}$ can ``hop'' to another column state $\ket{b}$ in the 2D problem,
instead focusing on how the columns interact with each other.

The essential physics
of the nematic state is captured by the following 1D Hamiltonian,
\begin{align*}
    \hat{H}_{1D} &= \sum_i \left[ t \hat{T}_i - \delta \hat{D}_i \right] \\
    &+ \frac{1}{2} R_b^6 \sum_{i \neq j} \bigg[ V_{ij}^{aa} \hat{P}^a_i \hat{P}^a_j + 
    V_{ij}^{ab} \hat{P}^a_i \hat{P}^b_j +
    V_{ij}^{ac} \hat{P}^a_i \hat{P}^c_j \\
    &+ V_{ij}^{bb} \hat{P}^b_i \hat{P}^b_j + 
    V_{ij}^{bc} \hat{P}^b_i \hat{P}^c_j +
    V_{ij}^{cc} \hat{P}^c_i \hat{P}^c_j \bigg].
\end{align*}
The local operators for the model are constructed in the
basis of the column states $\ket{a}, \ket{b}$, $\ket{c}$, and are defined as,
\begin{gather}
    \hat{D} = \left( \begin{array}{ccc}
    3 & 0 & 0 \\
    0 & 3 & 0 \\
    0 & 0 & 3
    \end{array} \right), \nonumber \\
    \hat{T} = \left( \begin{array}{ccc}
    0 & 1 & 1 \\
    1 & 0 & 1 \\
    1 & 1 & 0
    \end{array} \right), \nonumber \\
    \hat{P}^a = \left( \begin{array}{ccc}
    1 & 0 & 0 \\
    0 & 0 & 0 \\
    0 & 0 & 0
    \end{array} \right), \nonumber \\
    \hat{P}^b = \left( \begin{array}{ccc}
    0 & 0 & 0 \\
    0 & 1 & 0 \\
    0 & 0 & 0
    \end{array} \right), \nonumber \\
    \hat{P}^c = \left( \begin{array}{ccc}
    0 & 0 & 0 \\
    0 & 0 & 0 \\
    0 & 0 & 1
    \end{array} \right). \nonumber
\end{gather}
The same Hamiltonian can also be written in terms of spin-1 operators and is conceptually straightforward. The first
summation in $\hat{H}_{1D}$ contains the local terms, 
where $\delta \hat{D}$ is a direct mapping of
the $\delta \hat{n}$ term in the original 2D Hamiltonian (up to a scalar), and $t\hat{T}$
encodes local hopping between the 3 different column states. Microscopically, the hopping emerges virtually from the $\hat{\sigma}_x$ term in the 2D Hamiltonian.  The second summation is a direct mapping of
the interaction terms in 2D to the new basis, where the various $V_{ij}$ matrices contain
the different long-range interaction matrix elements between different pairs of columns. For a faithful
mapping, we have $V^{aa} = V^{bb} = V^{cc}$ and $V^{ab} = V^{ac} = V^{bc}$, and all of them
scale as $V_{ij} \sim \frac{1}{|i-j|^6}$.

\begin{figure}
    \centering
    \begin{tabular}{cc}
        \includegraphics[width=0.3\linewidth]{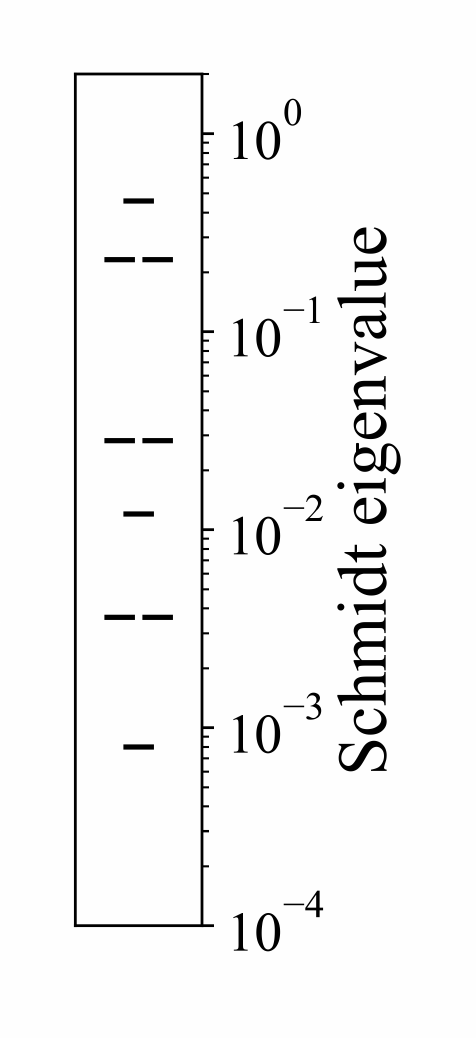} &
        \includegraphics[width=0.3\linewidth]{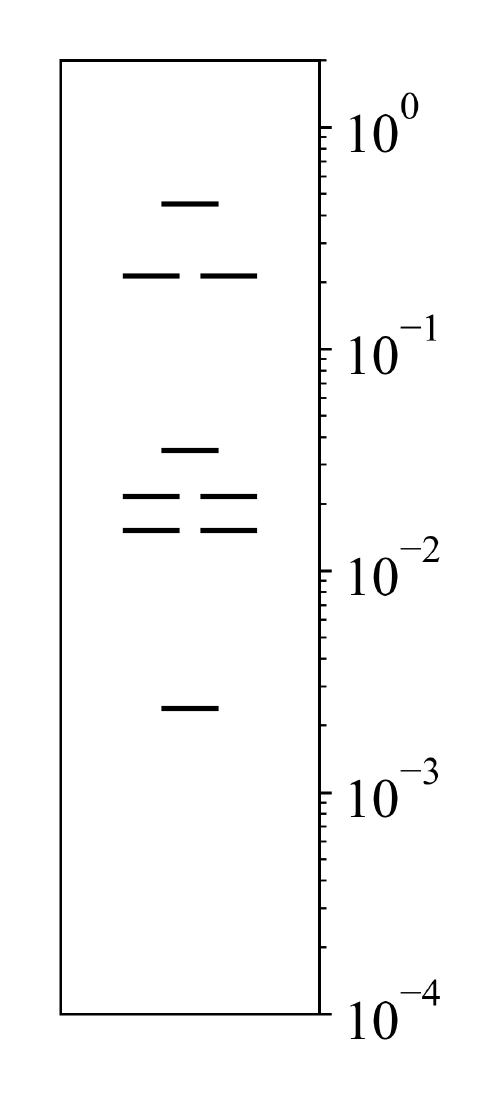}
    \end{tabular}
    \caption{The inter-column entanglement spectrum of the 2D nematic phase (left) and the 
    ground state of the 1D effective model presented in section~\ref{sec:1dmodel} 
    (right) for hopping parameter $t=0.5$. They both contain 3
    dominant eigenvalues with one doubly degenerate pair. The subsequent levels consist of
    two double degeneracies and two single degeneracies in both cases, although the spacing and
    ordering is slightly different.}
    \label{fig:1d_model_EE}
\end{figure}

Using infinite DMRG~\cite{mcculloch2008infinite}, we can obtain the ground state of this 1D model
in the thermodynamic limit. For $(\delta, R_b, t) = (5.0, 2.3, 0)$ we find that the ground state
is the unentangled crystal $\ket{abcabc...}$, as expected since this corresponds to the the classical
ground state in 2D and $t=0$ is the classical limit of this model. For $t > 0$ we find an entangled
ground state with equal density in the local $\ket{a}, \ket{b}$, and $\ket{c}$ basis states, as
in the 2D nematic ground state. We also observe that the structure of the entanglement spectrum
in the 1D model ground state is very similar to the entanglement spectrum of the 2D
nematic state in between the columns, 
as shown in Fig.~\ref{fig:1d_model_EE}. 
We conclude that this 1D effective Hamiltonian provides a useful model of the 2D nematic state. 

\section{Finite phase diagram: $15 \times 15$ and $16 \times 16$}
\label{sec:finite_details}

The phase diagram of the $15 \times 15$ lattice reported in the main text contained
many of the ground state orders seen in the bulk phase diagram, but it also revealed the
strong finite-size effects induced by the boundary. Due to the long-range van der Waals
interactions, Rydberg excitations at the edge of the array incur
roughly half of the energetic penalty that excitations in the interior do, but lower
the energy by an equal amount ($\delta$).
Except at small values of $R_b$, this induces excitations along the edge of the array
to be more densely packed than what would be expected from the bulk phase diagram at a
given point $(\delta, R_b)$. This generic effect causes frustration between the boundary
and interior of the finite lattices, which gives rise to the square classical order and many
defect-dominated states at large $R_b$, as discussed in the main text.
In these defect states, the optimal bulk density becomes so small relative to the optimal 
edge density that the ground states are permeated by edge-induced defects, leaving only
small regions of any discernible order and making the precise configuration 
very sensitive to small changes in $R_b$ and $\delta$.

In addition to the $15 \times 15$ lattice, we also studied two slices ($\delta = 4.0, 5.0$) 
of the phase diagram of the $16 \times 16$ lattice to probe for bulk-like ordered phases
where the $15 \times 15$ system is dominated by defects.
Specifically, we focused on the $R_b > 1.8$ region, for which the results
are shown in Fig.~\ref{fig:finite_pdiag}b.
We find a clear region of the stability for
the boundary-bulk frustrated $\frac{1}{5}$-stagger phase (red), for which the density profile
is shown in Fig. 4c of the main text. Along with a small region of the 3-star phase
(green and black), these
regions are unique to the $16 \times 16$ lattice (i.e. they are not seen in $15 \times 15$).
There are also some common features between the two array sizes, namely regions of the 
star and $\frac{1}{8}$-stagger (gold) phase as well as many defect states. This
suggests that the defect states are an intrinsic part of the physics of medium-sized arrays.

As reflected by the triangular markers in Fig.~\ref{fig:finite_pdiag}b (which reflect inconsistent convergence) we found it more challenging than the $15 \times 15$ 
lattice to systematically converge the PEPS calculations with respect to $(D, \chi)$, especially
in the star phase (blue). In part, this was due to the boundary itself being frustrated; on an
even-sided lattice it is not possible to place excitations in all corners and also along all edges
spaced by a distance of 2. Because the corner excitations are strongly pinned due to their reduced
interaction penalty, this causes the boundary to be frustrated and makes it more difficult to
prepare a good initial guess with our rudimentary strategies.

\section{Comparing to experiment: $9 \times 9$ and $13 \times 13$ lattices}
\label{sec:star_v_square}

\begin{figure}
    \centering
    %\includesvg[width=\linewidth]{figs/supp/expt_compare_supp.svg}
    \includegraphics[width=\linewidth]{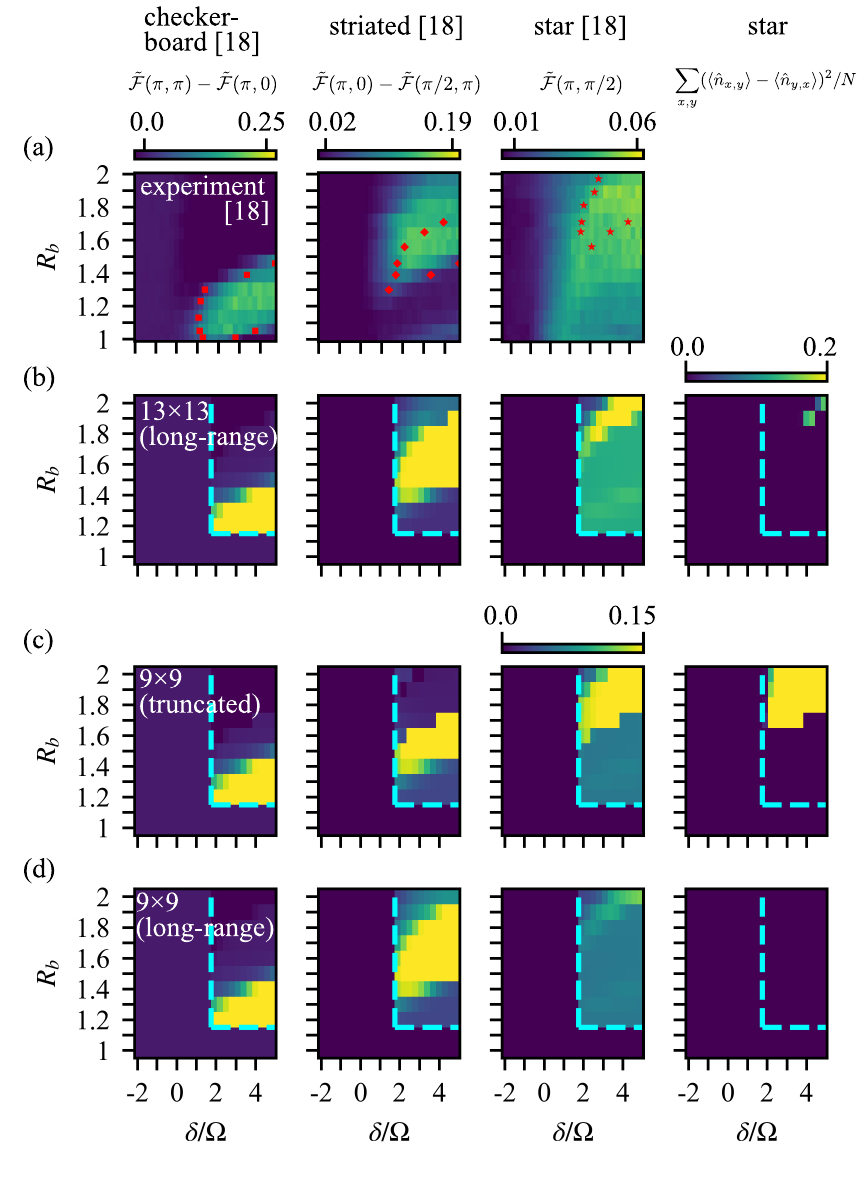}
    \caption{Detailed comparison to experimental phase diagram. 
    The (a) row directly reproduces the experimental phase diagram on the
    $13 \times 13$ lattice (data extracted from Ref.~\cite{ebadi2020} Fig. 4). 
    Rows (b)-(d) show analogous
    numerical data on $9 \times 9$ and $13 \times 13$ lattices, where (b) and (d)
    are results from simulations containing all long-range interactions and (c)
    shows results using interactions truncated to zero beyond distance 2. This
    is identical to the truncation scheme used in numerics in Ref.~\cite{ebadi2020}.
    The first three columns show all three order parameters used in~\cite{ebadi2020}
    to distinguish the phase diagram, while the fourth column shows a new, more precise
    order parameter for the star phase.
    Red dots in (a) denote the phase boundaries assigned in~\cite{ebadi2020}, while the 
    cyan dotted lines in (b)-(d) indicate the subset of parameter space that was computed.
    }
    \label{fig:expt_compare_supp}
\end{figure}

\begin{figure}
    \centering
    \includegraphics[width=\linewidth]{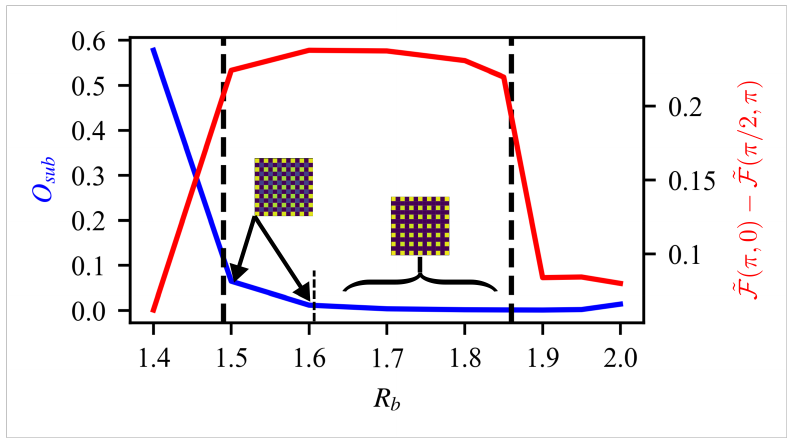}
    \caption{Distinguishing the striated and square orders on the $13 \times 13$ lattice
    at the slice $\delta=4.0$. The striated order parameter 
    $\tilde{\mathcal{F}}(\pi, 0) - \tilde{\mathcal{F}}(\pi/2, \pi)$ (red) is large
    across the range $R_b=1.5 - 1.85$, but the density of quantum fluctuations
    on the $(1,1)$-sublattice $O_{sub}$ (blue) decays to $\sim 0$ by 
    $R_b = 1.65$, revealing the square order.
    }
    \label{fig:expt_compare_supp2}
\end{figure}

The main text discussed discrepancies between
our numerical results on the $13 \times 13$ lattice and analysis reported in a recent experiment~\cite{ebadi2020}, specifically concerning the
striated, square and star phases. It was noted that the actual experimental data appears to agree with our numerics, but the interpretation of the data offered in Ref.~\cite{ebadi2020} is inconsistent with ours. 
This section details the effect of the approximations made in the numerics of Ref.~\cite{ebadi2020} on the interpretation of the data, and how relaxing those approximation leads to the interpretation described in our main text.

\subsection{Context}
In Ref.~\cite{ebadi2020}, the experimental data on the $13 \times 13$ square lattice was primarily
understood with respect to DMRG calculations performed on the $9 \times 9$ lattice
(all open boundaries), in which interactions were truncated to zero beyond a distance of 2. 
The experimental results of Ref.~\cite{ebadi2020} are reproduced in
Fig.~\ref{fig:expt_compare_supp}a,
and they are compared to our numerical results on $9 \times 9$ and $13 \times 13$ lattices 
(Fig.~\ref{fig:expt_compare_supp}b-d).
The region of the phase diagram that was studied included domains of stability for the
disordered, checkerboard, striated, and star phases. The square ``phase'' was not separately reported, although it may be considered the classical limit of the striated phase.

We also introduce here a useful order parameter for detecting the star phase,
\begin{equation}
    O_{star} = \sum_{x,y} (\langle \hat{n}_{x,y} \rangle - \langle \hat{n}_{y,x} \rangle)^2 
    / N,
\end{equation}
where $N = L_x \cdot L_y$.
$O_{star}$ detects a symmetry breaking that occurs in the star phase but not
in the disordered, checkerboard, striated, or square phases. 
On a finite lattice, this provides a clean way to define
the star phase separate from the other orders in this set.
We also recapitulate the definition of the order parameters defined in~\cite{ebadi2020}
and used in Fig.~\ref{fig:expt_compare_supp},
\begin{align}
    \tilde{\mathcal{F}}(k_1, k_2) &= (\mathcal{F}(k_1,k_2) + \mathcal{F}(k_2,k_1)) / 2 \\
    \mathcal{F}(k_1,k_2) &= | \sum_{x,y} \exp(i (k_1 x + k_2 y)) \langle \hat{n}_{x,y} \rangle| / N.
\end{align}

\subsection{Star phase stability}
In Fig.~\ref{fig:expt_compare_supp}c, we recompute the main $9 \times 9$ phase diagram numerical
results used in~\cite{ebadi2020}, which use truncated interactions. The bright region in 
$\tilde{\mathcal{F}}(\pi, \pi/2)$ predicts a large domain of stability for the star phase, which
is corroborated by the value of $O_{star}$. This data was used in~\cite{ebadi2020} 
to draw the expected
phase boundary in the $13 \times 13$ experimental data seen in Fig~\ref{fig:expt_compare_supp}a.
However, Fig.~\ref{fig:expt_compare_supp}d shows the analogous results on the $9 \times 9$
lattice when including all long-range interactions. Surprisingly,
the star phase gets completely destabilized! This illustrates the hazard
of interpreting the experimental data from smaller lattice simulations.

Unlike the $9\times 9$ lattice, we observe that the $13 \times 13$ lattice phase diagram has a qualitative difference: it hosts a nonzero domain of star phase even when accounting for all long-range interactions. As pointed
out in the main text, $\tilde{\mathcal{F}}(\pi, \pi/2)$ is not a sensitive order parameter for the star phase
as it appears on finite lattices, but $O_{star}$ does reveal the tiny stable region of the star phase (see Fig.~\ref{fig:expt_compare_supp}b).

\subsection{Square and striated phases}
The overestimation of the extent of the star phase 
by using numerics from the $9 \times 9$
lattice with truncated interactions also results in an underestimation of the
extent of the striated order parameter, 
$\tilde{\mathcal{F}}(\pi, 0) - \tilde{\mathcal{F}}(\pi/2, \pi)$,
since $\tilde{\mathcal{F}}(\pi/2, \pi)$ is the star order parameter used in~\cite{ebadi2020}
(see Fig~\ref{fig:expt_compare_supp}c). 
These $9 \times 9$ results were
used in~\cite{ebadi2020} to interpret the striated phase domain in the experimental data, 
so the boundary drawn in 
Fig.~\ref{fig:expt_compare_supp}a is too small. In fact, the extent of the
experimental data for $\tilde{\mathcal{F}}(\pi, 0) - \tilde{\mathcal{F}}(\pi/2, \pi)$
(Fig.~\ref{fig:expt_compare_supp}a) is significantly larger than the drawn boundary,
corresponding much more closely to the numerical data on the $13 \times 13$ including long-range 
interactions (Fig.~\ref{fig:expt_compare_supp}b), as mentioned in the main text.

In this work, we distinguish a region of classical square order from the striated phase 
where the square order
contains (almost) no quantum fluctuations on the $(1,1)$-sublattice, 
which are an essential feature of the striated phase in the bulk. 
$\tilde{\mathcal{F}}(\pi, 0) - \tilde{\mathcal{F}}(\pi/2, \pi)$ does not help distinguish between
square and striated orders, and no classical square order was reported in 
Ref.~\cite{ebadi2020}.
In Fig.~\ref{fig:expt_compare_supp2} 
we show that a large
part of the bright region in $\tilde{\mathcal{F}}(\pi, 0) - \tilde{\mathcal{F}}(\pi/2, \pi)$
on the $13 \times 13$ lattice should be interpreted as a classical square order by plotting,
\begin{equation}
    O_{sub} = \begin{cases} 
      \frac{4}{N} \sum_{x,y}\langle n_{x,y} \rangle  & \mathrm{if } x \bmod 2 =1, y \bmod 2 = 1 \\
      0 & \mathrm{else} \nonumber
   \end{cases}
\end{equation}
which detects the deformation of the density on the $(1,1)$-sublattice.
This sublattice is defined in terms of a $2 \times 2$ cell, as in~\cite{ebadi2020}.

\subsection{Numerical accuracy}
All numerical results in Figs.\ref{fig:expt_compare_supp}-\ref{fig:expt_compare_supp2}
were computed using DMRG. It was possible to study
the $13 \times 13$ lattice using DMRG because we only investigated a low-entanglement region of the phase diagram. The level of accuracy for these calculations is shown 
in Fig.~\ref{fig:dmrg_err_9x9} in terms of the 
largest truncated singular value during the DMRG sweep.
In the ordered regions of the results, the largest
truncated singular value is below $10^{-9}$, which is generally considered  accurate.

%%%%%%%%%%%%%%%%%%%%%%%%%%%%%%%%%%%%%%%%%%%%%%%%%%%%%%%%%%%%%%%%%%%%%%%%%%%%%%%%%%%%%%%%
% include the below on local machine, but comment out for submission
%\bibliographystyle{apsrev4-1}
%\bibliography{refs}

% include the below for submission, make sure above is commented out
%\input{main.bbl}

%\end{document}

%\input{supp_insert.bbl}

\end{document}